\documentclass[sigconf]{acmart}
\usepackage{color,soul}
\soulregister\cite7
\soulregister\ref7

\usepackage[table]{xcolor}
\usepackage{makecell}
\usepackage{pifont}
\usepackage{xspace}

\usepackage{amsmath,amsfonts}
\usepackage{algorithm}
\usepackage{algorithmic}

\usepackage{textcomp}

\usepackage{lipsum} 
\usepackage{booktabs} 
\usepackage{multirow} 
\usepackage{comment}
\usepackage{graphicx}
\usepackage{caption}
\usepackage{subcaption}
\usepackage{arydshln}

\newcommand{\squishlist}{
 \begin{list}{$\bullet$}
  { \setlength{\itemsep}{0pt}
     \setlength{\parsep}{0pt}
     \setlength{\topsep}{3pt}
     \setlength{\partopsep}{0pt}
     \setlength{\leftmargin}{1.5em}
     \setlength{\labelwidth}{1em}
     \setlength{\labelsep}{0.5em} } }

\newcommand{\squishnums}{
 \begin{list}{$\bullets$}
  { \setlength{\itemsep}{0pt}
     \setlength{\parsep}{3pt}
     \setlength{\topsep}{3pt}
     \setlength{\partopsep}{0pt}
     \setlength{\leftmargin}{1.5em}
     \setlength{\labelwidth}{1em}
     \setlength{\labelsep}{0.5em} } }

\newcommand{\squishlisttwo}{
 \begin{list}{$\bullet$}
  { \setlength{\itemsep}{0pt}
     \setlength{\parsep}{0pt}
    \setlength{\topsep}{0pt}
    \setlength{\partopsep}{0pt}
    \setlength{\leftmargin}{2em}
    \setlength{\labelwidth}{1.5em}
    \setlength{\labelsep}{0.5em} } }

\newcommand{\squishend}{
  \end{list}  }

\newif\ifcommenton
\commentontrue

\ifcommenton
\newcommand{\JT}[1]{{\color{brown}\bfseries [Jianming: #1]}}
\newcommand{\YC}[1]{{\color{orange}\bfseries [Yangyu: #1]}}
\newcommand{\AB}[1]{{\color{olive}\bfseries [Abhi: #1]}}
\newcommand{\alexey}[1]{\textcolor{navy}{[Alexey: #1]}}
\newcommand{\alind}[1]{\textcolor{codegreen}{[Alind: #1]}}
\newcommand{\DP}[1]{{\color{green}\bfseries [David: #1]}}
\newcommand{\HEO}[1]{{\color{blue}\bfseries [Taekyung: #1]}}
\newcommand{\fixme}[1]{{{\color{blue} #1}}}
\else
\newcommand{\JT}[1]{}
\newcommand{\YC}[1]{}
\newcommand{\AB}[1]{}
\newcommand{\alexey}[1]{}
\newcommand{\alind}[1]{}
\newcommand{\DP}
\newcommand{\HEO}[1]{}
\newcommand{\fixme}
\fi

\definecolor{codegreen}{rgb}{0,0.6,0}
\definecolor{codegray}{rgb}{0.5,0.5,0.5}
\definecolor{codepurple}{rgb}{0.58,0,0.82}
\definecolor{backcolour}{rgb}{0.95,0.95,0.92}
\definecolor{purple}{RGB}{128,0,128}
\definecolor{indigo}{RGB}{75,0,130}
\definecolor{royalblue}{RGB}{65,105,225}
\definecolor{navy}{RGB}{0,0,128}
\definecolor{myred}{RGB}{153,0,0}
\definecolor{myblue}{RGB}{0,76,153}
\definecolor{mygreen}{RGB}{0,102,51}
\definecolor{mypurple}{RGB}{76,0,153}

\newcommand{\secref}[1]{\S\ref{#1}}
\newcommand{\figref}[1]{Fig.~\ref{#1}}
\newcommand{\tabref}[1]{Tab.~\ref{#1}}
\newcommand{\algref}[1]{Alg.~\ref{#1}}

\newcommand{\frameworknospace}{\textsc{MORPH}\xspace}%
\newcommand{\framework}{\textsc{MORPH} }%

\newcommand{\Fm}[1][M]{\ensuremath{\color{myblue}{\mathbb{F}_{#1}}} }
\newcommand{\Bm}[1][M]{\ensuremath{\color{myblue}{M}} }

\newcommand{\Fmxs}[1][M]{\ensuremath{\color{myblue}{\mathbb{F}_{#1}}}\xspace}
\newcommand{\Bmxs}[1][M]{\ensuremath{\color{myblue}{M}}\xspace}

\newcommand{\Fqxs}[1][Q]{\ensuremath{\color{myred}{\mathbb{F}_{#1}}}\xspace}
\newcommand{\Rqxs}{\ensuremath{\color{myred}{Q}}\xspace}

\newcommand{\Fpxs}[1][P]{\ensuremath{\color{mygreen}{\mathbb{F}_{#1}}}\xspace}
\newcommand{\Gp}{\ensuremath{\color{mygreen}{P}}\xspace}
\newcommand{\Gpxs}{\ensuremath{\color{mygreen}{P}}\xspace}
\author{Jianming Tong}
\authornote{Both authors contributed equally to this research.}
\email{jianming.tong@gatech.edu}
\affiliation{%
  \institution{Georgia Institute of Technology}
  \city{Atlanta}
  \state{Georgia}
  \country{USA}
  \postcode{30332}
}

\author{Jingtian Dang}
\authornotemark[1]
\email{dangjingtian@gatech.edu}
\affiliation{
  \institution{Georgia Institute of Technology}
 \city{Atlanta}
 \state{Georgia}
 \country{USA}}

\author{Simon Langowski}
\email{slangowski@mit.edu}
\affiliation{%
 \institution{Massachusetts Institute of Technology}
  \city{Cambridge}
  \state{Massachusetts}
  \country{USA}}

\author{Tianhao Huang}
\email{tianhaoh@mit.edu}
\affiliation{%
  \institution{Massachusetts Institute of Technology}
  \city{Cambridge}
  \state{Massachusetts}
  \country{USA}
}

\author{Asra Ali}
\email{asraa@google.com}
\affiliation{%
 \institution{Google}
 \city{Austin}
 \state{Texas}
 \country{USA}}

\author{Jeremy Kun}
\email{jkun@google.com}
\affiliation{%
 \institution{Google}
 \city{Portland}
 \state{Oregon}
 \country{USA}}

\author{Jevin Jiang}
\email{jevinjiang@google.com}
\affiliation{%
 \institution{Google}
 \city{Sunnyvale}
 \state{California}
 \country{USA}}

\author{Srinivas Devadas}
\email{devadas@mit.edu}
\affiliation{%
 \institution{Massachusetts Institute of Technology}
  \city{Cambridge}
  \state{Massachusetts}
  \country{USA}}

\author{Tushar Krishna}
\email{tushar@ece.gatech.edu}
\affiliation{%
  \institution{Georgia Institute of Technology}
  \city{Atlanta}
  \state{Georgia}
  \country{USA}
  \postcode{30332}}

\copyrightyear{2026}
\acmYear{2026}
\setcopyright{cc}
\setcctype{by}
\acmConference[DAC '26]{63rd ACM/IEEE Design Automation Conference}{July 26--29, 2026}{Long Beach, CA, USA}
\acmBooktitle{63rd ACM/IEEE Design Automation Conference (DAC '26), July 26--29, 2026, Long Beach, CA, USA}
\acmDOI{10.1145/3770743.3804402}
\acmISBN{979-8-4007-2254-7/2026/07}

\begin{document}

\title{Enabling AI ASICs for Zero Knowledge Proof}

\begin{abstract}
\label{sec:abstract}
Zero-knowledge proof (ZKP) provers remain costly because multi-scalar multiplication (MSM) and number-theoretic transforms (NTTs) dominate runtime as they need significant computation. AI ASICs such as TPUs provide massive matrix throughput and SotA energy efficiency. 
We present \emph{MORPH}, the first framework that reformulates ZKP kernels to match AI-ASIC execution. We introduce \emph{Big-$T$} complexity, a hardware-aware complexity model that exposes heterogeneous bottlenecks and layout-transformation costs ignored by Big-$O$. Guided by this analysis, (1) at arithmetic level, MORPH develops an MXU-centric extended-RNS lazy reduction that converts high-precision modular arithmetic into dense low-precision GEMMs, eliminating all carry chains, and (2) at dataflow level, MORPH constructs a unified-sharding layout-stationary TPU  
Pippenger MSM and optimized 3/5-step NTT that avoid on-TPU shuffles to minimize costly memory reorganization. 
Implemented in JAX, MORPH enables TPUv6e8 to achieve up-to 10$\times$ higher throughput on NTT and comparable throughput on MSM than GZKP. Our code: \url{https://github.com/EfficientPPML/MORPH}.
\end{abstract}

\maketitle
\thispagestyle{plain}
\pagestyle{plain}

\vspace{-2mm}
\section{Introduction}
\label{sec:introduction}

Zero-knowledge proofs (ZKPs) have evolved from a theoretical construct into a core building block for practical verifiable computation and scalable blockchains. They allow services to outsource computation while keeping data private and verification cheap, but the\textbf{ prover remains prohibitively expensive}, e.g., generating a proof for ImageNet ViT requires nearly one hour~\cite{zhang2025zkvc}. Across widely deployed protocols~\cite{groth2016size,gabizon2019plonk}, prover runtime is dominated by \textbf{multi-scalar multiplication (MSM)} and the \textbf{number-theoretic transform (NTT)}, which commonly account for $\approx$70\% and 20–30\% of total latency, respectively~\cite{DistMSM,zhang2021pipezk}.

\begin{figure}[t]
    \centering
    \setlength{\fboxrule}{2pt}
    {\includegraphics[width=\columnwidth]{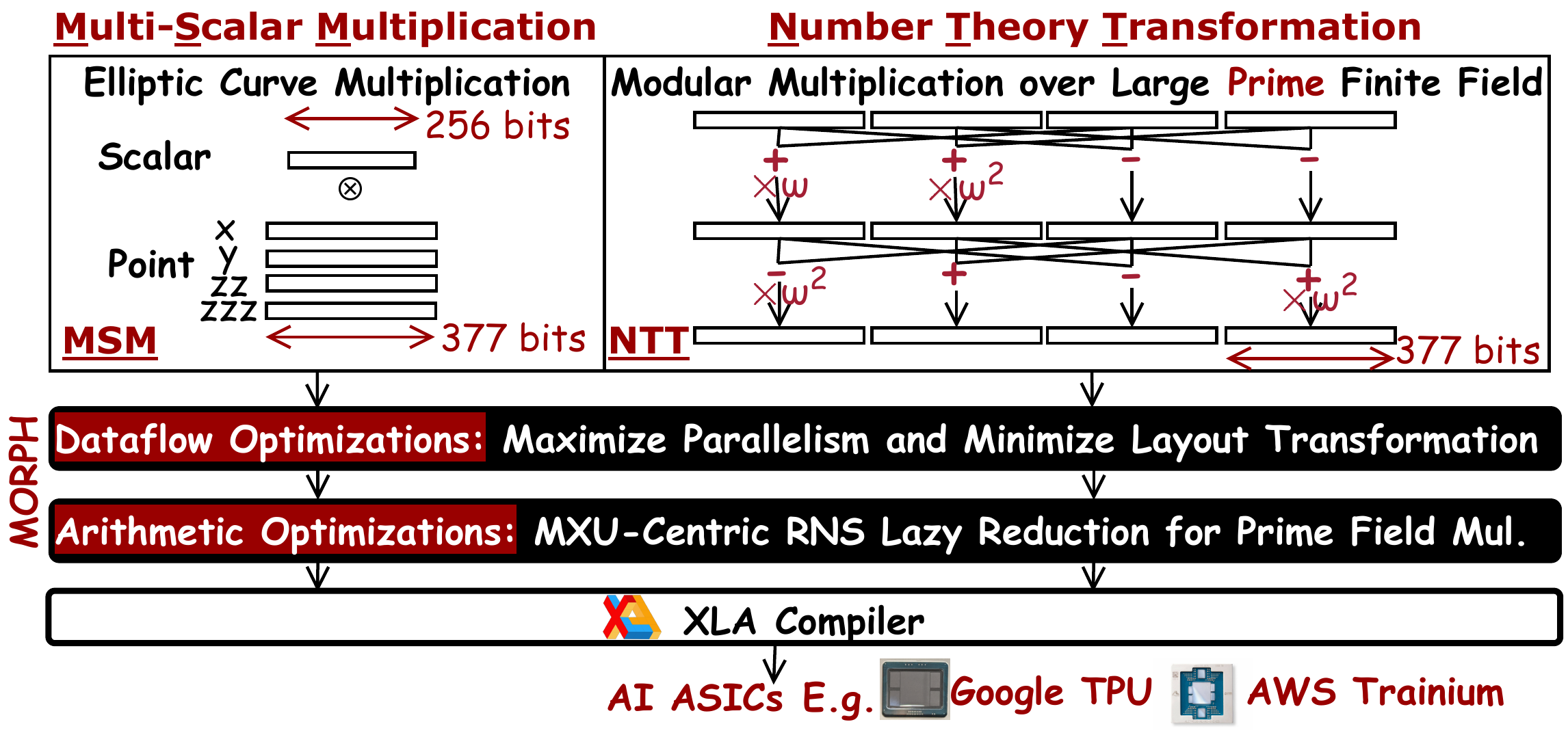}} 
    \vspace{-8.5mm}
    \caption{\frameworknospace's deployment flow with dataflow and arithmetic optimizations to accelerate ZKP on TPU.}
    \label{fig:prove_optimizations}
    \vspace{-6.9mm}
\end{figure}

Both MSM and NTT expose abundant parallelism at practical ZKP sizes, making parallel hardware promising for acceleration. Among these platforms, AI ASICs (Google TPU~\cite{jouppi2023tpuv4opticallyreconfigurable}, AWS Trainium~\cite{Trainium} etc.) offer extreme compute density and energy efficiency, significantly outperforming general-purpose accelerators at scale~\cite{11408507}. Architecturally, a TPU contains giant heterogeneous cores, including a matrix multiplication engine (MXU) and a vectorized processing engine (VPU). Each MXU/VPU contains orders of magnitude (1024/16$\times$) more multiply-accumulates (MACs) than one corresponding core in a GPU~\cite{scaling_book}. All MACs in a MXU/VPU execute instruction in the lock-step (SIMD) to substantially reduce per-operation control overhead and improve energy efficiency. 
This paper studies how to \textbf{systematically repurpose AI ASICs, particularly Google's TPUs~\cite{jouppi2023tpuv4opticallyreconfigurable}, for ZKP kernels (MSM/NTT)} and achieve higher energy efficiency and SoTA throughput at scale.

Existing ZKP acceleration spans GPUs~\cite{ma2023gzkp, DistMSM,cuHE, gzkp}, FPGAs~\cite{CycloneMSM, PriorMSM, HardcamlMSM, pottier2025optimsm, ohno2025acceleratingellipticcurvepoint, heax} and ASICs~\cite{nocap,zhang2021pipezk, UniZK}. While these systems hand-craft MSM/NTT algorithms for their respective platforms, they provide \textit{no principled way to quantify how far an MSM/NTT algorithm is from the platform's performance limits with detailed reasoning}. 
Some works provide Big-O complexity~\cite{sharp}, yet we find a surprising phenomenon: the fastest GPU/TPU algorithms intentionally use the algorithm with higher Big-O to expose massive parallelism and thus achieve higher throughput~\cite{11408507}. \textit{Big-O alone is insufficient for modern heterogeneous parallel hardware.}

Furthermore, Big-O cannot capture layout transformation cost, which becomes the dominant bottleneck on AI ASICs. When we port the SotA MSM and NTT algorithms for GPU/FPGA/ZKP-ASIC to TPUs with similar overall throughput, they slow down by 30$\times$ in our evaluation. This is because a TPU \textbf{\textit{only} operates on coarser-grained contiguous data (a 4KB vector register, VReg)}, and fine-grained data gathering in SotA MSM/NTT algorithms requires expensive shuffles and transposes to move data into contiguous VRegs. These layout-induced stalls are invisible under the traditional Big-O model. 
To address this gap, we introduce Big-$T$ notation, a platform-aware complexity model that measures the sequential bottleneck across heterogeneous pipelined compute units and the latency of layout transformation, providing an asymptotic lower bound for parallel execution on AI ASICs. Big-$T$ exposes two structural inefficiencies in SotA MSM/NTT algorithms on TPUs.

\textbf{Arithmetic-level Challenge}: ZKP requires modular arithmetic over large prime fields (e.g., 256/377/753 bits), but hardware natively supports much lower precision (e.g., 8/32-bit). Because prime fields lack a natural RNS factorization, the SotA algorithm~\cite{gzkp,yrrid_gpu_msm} performs high-precision arithmetic via digit decomposition and a chain of carry propagation, achieving <1\% compute utilization on TPUs.

\textbf{Dataflow-level Challenge}: The SotA dataflows incur prohibitive communication and storage overheads. In presorted Pippenger MSM~\cite{DistMSM,gzkp}, sorting is distributed across many GPU CUDA cores. Directly porting this strategy to TPU introduces prohibitive inter-TPU communication. Layout-invariant 3-step NTT~\cite{11408507} requires parameters that scale linearly with problem size, resulting in out-of-memory (OOM) failures at large scale.

In resolving these inefficiencies, we propose \textbf{\frameworknospace}\footnote{MORPH: \underline{M}atrix-\underline{O}riented \underline{R}eformulation of ZKP for \underline{P}arallel \underline{H}eterogeneous AI ASICs}, the \textit{first} framework enabling TPU as ZKP accelerator with arithmetic and dataflow level optimizations, as shown in \figref{fig:prove_optimizations}.

\textbf{Arithmetic-level Solution}: Since a prime field cannot be represented in a Residue Number System (RNS), \framework adopts~\cite{large_rns,SimonRNS} to encapsulate modular multiplication in a prime field in a \textit{larger non-prime finite field} that has a RNS representation. This enables high-precision multiplication to be expressed as independent 32-bit vector multiplications with no carry propagation. Then, \framework reforms the modular reduction (RNS reconstruction → modular reduction → RNS decomposition) as low-precision matrix multiplication and parallel vectorized operations to be accelerated by TPU’s MXU and VPU, achieving up-to 90$\times$ speedup.

\textbf{Dataflow-level Solution}: To tackle communication and layout reorganization, MORPH proposed \textbf{LS-PPG} to schedule communication-free workload dimensions across devices to avoid inter-TPU communication, ensures layout invariant within individual TPU. To mitigate parameters storage overflow, MORPH's \textbf{5-step NTT} recursively partitions workloads to reduce parameter size while preserving the same computational complexity. \label{sec:name_acry}

Together, \framework enables TPUv6e-8 to achieve up-to 10$\times$ NTT throughput and 1.2$\times$ MSM throughput than GZKP \cite{gzkp} (NVIDIA V100). \textbf{MORPH makes TPU the SotA throughput machine for high-precision NTT}, a promising platform for ZKP acceleration. Our contributions are:

\squishlist
\item \textbf{BIG-$T$ Complexity:} A hardware-aware complexity metric that captures heterogeneous compute spans and layout transformation, enabling principled analysis of MSM/NTT designs on AI ASICs and guiding the construction of layout-stationary pippenger for MSM and 5-step NTT, giving 5.8$\times$ and 1.6$\times$ speedup.
\item \textbf{MXU-centric RNS Lazy Reduction:} A matrix-oriented modular reduction for big integer multiplication in prime fields, which replaces sequential carry-propagation arithmetic with MXU-accelerated RNS-lazy reduction, achieving up-to $90\times$ speedup over the SotA radix decomposed Montgomery reduction.
\item \textbf{LS-PPG and 5-step Layout Invariant NTT:} Both avoid inter-TPU communication and minimize intra-TPU layout reorganization, delivering up-to 5.8$\times$ speedup over the SoTA MSM algorithm\cite{DistMSM} on TPU, and reduce parameters storage at scale. 
\squishend

\section{Background and Related Work}

\begin{figure}[t!]
    \centering
    \includegraphics[width=0.95\linewidth]{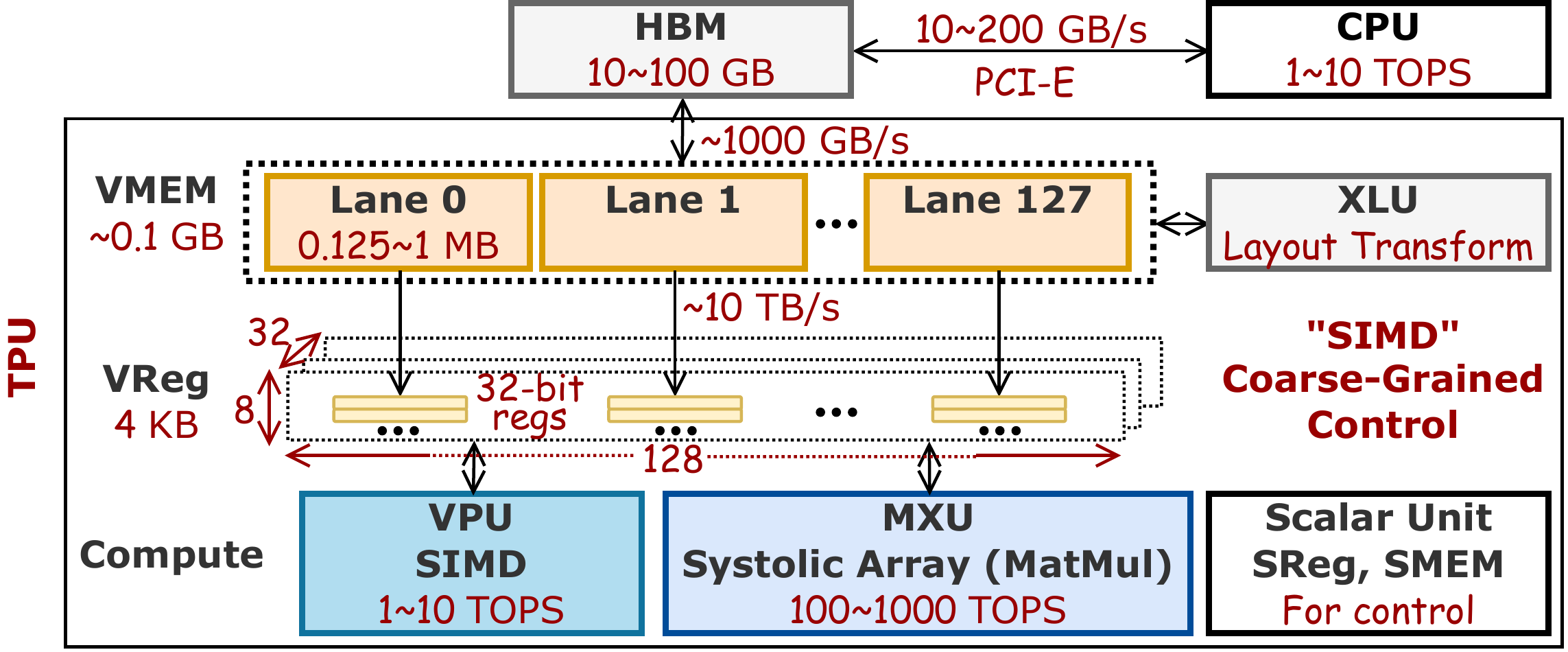}
    \resizebox{\columnwidth}{!}{
    \begin{tabular}{ccccc}\hline
    \textbf{} &\textbf{\underline{S}huffle} &\textbf{\underline{T}ranspose} &\textbf{MXU} &\textbf{VPU} \\\hline
    Max Parallelism &$PAR_{\text{S}}=4096$ &$PAR_{\text{T}}=4096$ &$PAR_{\text{MXU}}=131072$ &$PAR_{\text{VPU}}=2048$ \\
    Datatype &u32 &u32 &bf16/i8 &u32 \\
    Requirement & (8, 128) &(8, 128) &(8$\times$128$\times$128 MatMul) & 2048 \\
    \hline
    \end{tabular}}
    \vspace{-4mm}
    \caption{TPU programming model. All compute units operate on VRegs with $(8\times128)$ 32-bit registers each. Performance hinges on data being laid out such that tiles needed for computation can be loaded directly from VRegs. 
    Values from~\cite{jouppi2023tpuv4opticallyreconfigurable}.}
    \label{fig:tpu_programming_model}
    \vspace{-6.5mm}
\end{figure}

\subsection{Tensor Processing Unit (TPU)}
The TPU~\cite{jouppi2023tpuv4opticallyreconfigurable} is Google’s AI ASIC designed to accelerate machine learning training and inference with high energy efficiency. All on-chip computation is organized around the \textbf{Vector Register (VReg)} abstraction. Each VReg consists of \textbf{8$\times$128} of \textbf{32}-bit values executed in lock-step SIMD. All compute units on a TPU, including the VPU, MXU, and XLU, consume and produce data strictly at VReg granularity, as shown in \figref{fig:tpu_programming_model}. This uniform but coarse layout enables high efficiency but restricts flexibility: workloads whose shapes do not align with the fixed $(8,128)$ structure incur layout transformation because only full VRegs can be processed each cycle.

\noindent $\bullet$ \textbf{Vector Processing Unit (VPU).}
The VPU performs 32-bit SIMD arithmetic on VRegs. E.g., TPUv4 contains a dual-issue ALU to operate two VRegs concurrently, yielding a peak parallelism of $P_{\text{VPU}} = 2048$ 32-bit operations per cycle. Vectors with length not a multiple of (8, 128) cannot fully utilize VPU’s available parallelism. \\
\noindent $\bullet$ \textbf{Matrix Multiplication Unit (MXU).}
The MXU in TPUv4 is a large systolic array (128$\times$128) for int8 and bf16 matrix multiplication. It preloads weight tiles from VRegs and streams activation tiles from VRegs. When normalized to the same datatype, the MXU achieves roughly \textbf{16$\times$ higher peak throughput} than the VPU. Full efficiency requires workloads to conform to the MXU’s minimum effective tile of \textbf{8$\times$128$\times$128}. Larger gives more weights reuse while smaller or non-aligned matrix shapes leave MXU under-utilized. \\ 
\noindent $\bullet$ \textbf{Layout Transformation Unit (XLU).}
XLU performs shuffles, transposes, broadcasts, and reductions of VRegs. XLU is efficient for transformations at VRegs granularity, but costly for element-wise permutations. Shuffling $N$ individual elements may cost \textbf{$N$ cycles}. \\
\noindent $\bullet$ \textbf{HBM }and the \textbf{host CPU} serve as the off-chip data sources, but their bandwidth is typically \textbf{an order of magnitude lower} than on-chip VReg bandwidth. Consequently, high-performance TPU kernels favor \textbf{on-chip data reuse and layout transformation}.

\subsection{Performance Metric and Analysis} TPU’s heterogeneous compute units, each with different parallelism and tiling constraints, make classical models such as Big~$O$ and the work–span model inadequate. These models measure total computational work or critical-path latency under the assumption of a homogeneous sequential machine like a CPU, and therefore cannot capture VReg granularity, systolic-array tiling, or the cost of on-chip layout transformations. Further, Roofline analysis\cite{roofline_model} and trace-based profiling quantify how far execution is from hardware limits, but they still do not explain why performance is low. 

\subsection{Zero-knowledge Proof (ZKP) Acceleration} 

\subsubsection{Background on Kernel and Arithmetic in ZKP} \hfill

\textbf{\textit{Multi-Scalar Multiplication}}. MSM~\cite{groth2016size} computes $\sum_{n=0}^{N-1} S_n \circledast P_n$ 
where each scalar $S_n$ is $S^{BL}$ bits and each point $P_n$ lies on an elliptic curve that supports point addition (PADD, $\oplus$). A scalar–point multiplication ($\circledast$) is realized as repeated PADDs for $S_n$ times.

\textbf{\textit{Number Theory Transformation}}. NTT~\cite{groth2016size} converts a polynomial of degree $N$ into its frequency-domain representation by evaluating it at $N$ distinct roots of unity in a finite field. Given an input vector $\mathbf{x}$ of degree $N$, NTT computes $\textbf{X}_k=\sum_{n=0}^{N-1} x_n \times \omega_N^{kn} \bmod \Bmxs$, $k\in[0,N)$, where $\omega_N$ is a primitive $N$-th root of unity.

\textbf{\textit{Large prime field}}. Modern ZKP systems~\cite{groth2016size} rely on arithmetic over a finite field $\Fmxs$ with a \textbf{large} \textbf{prime} modulus $\Bmxs$ ($256/377/753$ bits). In both elliptic-curve MSM and polynomial-commitment NTT, all additions, multiplications are executed modulo large prime $\Bmxs$.

\vspace{-2mm}
\subsubsection{Related Work: Arithmetic Implementations} \hfill 

ZKPs operate over large prime fields $\Fm$ with $\log_2({\Bmxs})\ge 256$, exceeding native 32-bit word size on GPUs and TPUs. SotA GPU implementations~\cite{DistMSM, gzkp} therefore represent each field element as a vector of $D$ 32-bit ``digits''. A single field multiplication requires two stages:
(1) \textit{High-precision multiplication}, which performs $O(D^2)$ 32-bit multiplications followed by a $O(D)$ sequential carry propagation of length $D$; and
(2) \textit{Montgomery reduction}, which reduces the $2D$-digit intermediate product back to $D$ digits, incurring another $O(D^2)$ 32-bit multiplications. We use this SotA radix Montgomery reduction as our \textbf{arithmetic baseline}, termed as \textbf{\textit{radix Mont}}.

\textbf{RNS} represents an integer $x$ by residues $(x \bmod q_i)$ over coprime moduli $\{q_i\}$, where modulus ${\Rqxs}\!=\!\prod_i q_i$. RNS could lower $O(D^2)$ down to $O(D)$ when performing modular multiplication with respect to $\Rqxs$. But the prime modulus $\Bmxs$ cannot be factored into multiple residues, making such reduction inapplicable to ZKP.

\vspace{-2mm}
\subsubsection{Related Work: Dataflow Optimizations} \hfill
\label{sec:baseline}
\textbf{Presorting Pippenger (PPG) - MSM} In prior work of MSM acceleration on GPU \cite{ma2023gzkp, DistMSM}, FPGA \cite{CycloneMSM,PriorMSM,HardcamlMSM, pottier2025optimsm, ohno2025acceleratingellipticcurvepoint} and ASICs \cite{zhang2021pipezk, UniZK, SZKP, legoZK, gypsophila}, Presorting PPG is the SotA MSM algorithm. Its key idea is to \emph{reduce the total number of point additions} by first grouping points whose scalars share the same value, summing those points once, and then multiplying the accumulated result by that scalar value. To increase the likelihood of such sharing (``collisions''), each scalar is split into $K = \left\lceil {S^{BL}}/{c} \right\rceil$ windows of $c$ bits. We adopt presorting PPG as our \textbf{dataflow baseline}, termed as \textbf{``Presort-PPG''}. 

\textit{\textbf{Butterfly NTT} and \textbf{layout invariant 3-step NTT}}.
SotA NTT acceleration on CPUs \cite{seal, intelhexl}, GPUs \cite{cuHE, gzkp}, FPGAs \cite{heax} and ASICs \cite{f1he} implements recursive Cooley-Tukey NTT with minimal computation complexity $O(NlogN)$, commonly called as ``butterfly NTT" in \figref{fig:ntt_butterfly}. However, each stage performs fine-grained shuffles that are smaller than vector-register width, leading to costly layout transformations, making TPUs inefficient. Recent TPU work~\cite{11408507} proposes a \textit{layout-invariant 3-step NTT} achieving $O(N^{3/2})$ arithmetic with zero layout transformation, breaking the NTT throughput record of the butterfly NTT on GPUs~\cite{fan2022tensorfheachievingpracticalcomputation, kim2024cheddar} on lower-precision ${\Bmxs}<32\rm\ bits$ and lower degree ($N\leq2^{17}$). It reforms an $N$-input NTT into an $(R,C)$ matrix ($N=R\times C$) such that an NTT is reformed into matrix and vectorized multiplication for TPU acceleration, as shown in \figref{fig:three_step_ntt}. \textbf{We adopt 3-step NTT as dataflow baseline.}

\section{Method}
This section first defines the Big $T$ notation, then uses it to analyze arithmetic and dataflow inefficiencies of the SotA MSM/NTT algorithm on TPUs, finally showing how \frameworknospace resolves them. 



\vspace{-2mm}
\subsection{Big-$T$ Complexity and TPU's Parallelism}
Today's parallel computing devices, including GPUs and TPUs, often consist a heterogeneous set of pipelined compute units $\mathcal{U} = \{U_1, U_2, \dots, U_K\}$,
where each $U_k$ denotes a class of units. Let $P_k$ be the number of parallel instances of unit type $U_k$. We decompose the total work of size $N$ into per-unit contributions $W_k(N), k\in [1,\dots,K]$, where $W_k(N)$ counts the operations mapped to unit $U_k$.

All these on-chip compute units are deeply pipelined, the execution time is bounded by the slowest (bottleneck) pipeline stage. We define the \emph{BIG-$T$ complexity} (termed as Big-$T$ notation) of an algorithm on a parallel computing device as
\[
  T(N) = O\Bigl(
    \max\bigl\{
      \max_{k} \tfrac{W_k}{P_k},
      \; Mem
    \bigr\}
  \Bigr)
\]
if there exists a constant $c > 0$ and $N_0$ such that for all $N \ge N_0$,
$
  T(N) \le
  c \cdot
  \max\Bigl\{
    \max_{k} \frac{W_k}{P_k},
    \; Mem
  \Bigr\}.
$
Here, the term ``$\max_k 
\frac{W_k}{P_k}$'' captures the bottleneck among heterogeneous pipelined compute units. ``$Mem$'' captures the overall latency of off-chip data access. \textbf{Together, they provide an asymptotic lower latency bound of a proposed algorithm on parallel computing devices.} Under sufficiently large workload, the ideal parallelism of compute units in a TPU used for Big-$T$ analysis is listed in \figref{fig:tpu_programming_model}. 



\vspace{-2mm}
\subsection{Arithmetic Optimizations}
\begin{table}[!htp]\centering
\vspace{-5mm}
\caption{Big-$T$ Complexity of Arithmetic (Bottleneck in {\color{myred}{Red}})}\label{tab:arithmetic_bigt}
\vspace{-4mm}
\resizebox{\columnwidth}{!}{
\begin{tabular}{c|cccc}\hline
\textbf{Kernel} & \textbf{VPU Span} & \textbf{MXU Span} & \textbf{XLU Span} & \textbf{Memory Span} \\\hline
\rowcolor[gray]{0.9}
Radix Mont.
  & $\displaystyle \frac{D^{2}}{PAR_{\text{VPU}}}$ 
  & $\displaystyle \frac{D^{2}}{PAR_{\text{MXU}}}$ 
  & {\color{myred}{$\displaystyle \frac{D^{2}\log D}{PAR_{\text{S}}}$}}
  & $\displaystyle \frac{D}{BW_{\text{HBM}}}$ \\
\rowcolor[HTML]{E6EFDB}
MXU RNS Lazy
  & {\color{myred}{$\displaystyle \frac{4D}{PAR_{\text{VPU}}}$}} 
  & $\displaystyle \frac{D^{2}}{PAR_{\text{MXU}}}$ 
  & $\sim$0
  & $\displaystyle \frac{2D}{BW_{\text{HBM}}}$ \\ \hline
\end{tabular}}
\vspace{-6mm}
\end{table}

\subsubsection{Challenge: Radix Montgomery Reduction is Dominated by Memory Reorganization} For a value with $D$ 32-bit digits, radix-$2^{32}$ Montgomery multiplication requires
(1) a $O(D^2)$ big-integer multiplication and
(2) a $O(D^2)$ Montgomery reduction,
each containing a sequential $D$-step carry-propagation chain.
Carry propagation demands fine-grained shuffling and index permutation on a TPU, which appears as the XLU span bottleneck in \tabref{tab:arithmetic_bigt}.
Thus, the limiting factor is not compute throughput, but the repeated memory reorganization inherent in Radix Montgomery Reduction.

\subsubsection{Solution: Reducing $O(D^2)\rightarrow O(2D)$ for Big-Integer Multiplication via RNS} 
To eliminate quadratic-cost digit multiplications, we encode each high-precision prime-field element in an extended non-prime field $\Fqxs$ following \cite{large_rns}.
We select $\Rqxs > \Bmxs^2$ to guarantee that for any $x,y\in\Fqxs$, the product of their RNS vectors never overflows $\Rqxs$, avoiding a true reduction by $\Rqxs$.
Hence, the multiplication is fully decomposed into independent 32-bit limb multiplications. \textit{This transformation converts the compute pattern from quadratic all-to-all digit multiplications into linear digit-independent multiplications}.

\subsubsection{Solution: Reducing $O(D^2)$ Reduction via \textbf{MXU-Centric RNS Lazy Reduction}} 
Although multiplication is linearized in $\Fqxs$, modular reduction by $\Bm$ must still be performed over the prime field $\Fmxs$.
A naïve approach would require RNS reconstruction from $\Fqxs$ $\rightarrow$ prime-field reduction ($\bmod\ {\Bmxs})$ $\rightarrow$ RNS decomposition into $\Fqxs$.

We reduce this overhead by applying Basis Aligned Transformation~\cite{11408507} to Simon’s reduction~\cite{SimonRNS}, collapsing this multi-stage pipeline into one 8-bit matrix multiplication (Lines 15 and 18 in Alg.\ref{alg:mxu_centric_modmul}) plus four 32-bit vector operations (Lines 16,17,19,20 in Alg.\ref{alg:mxu_centric_modmul}). This shifts the dominant $O(D^2)$ term to a low-precision matrix multiplication, with its cost amortized by MXU, yielding $O(D)$ latency overhead. It also removes all carry-propagation from the critical path, making it throughput-bound rather than shuffle-bound.

\subsubsection{Walk-Through Example}
\begin{figure}[t!]
    \centering
    \includegraphics[width=\linewidth]{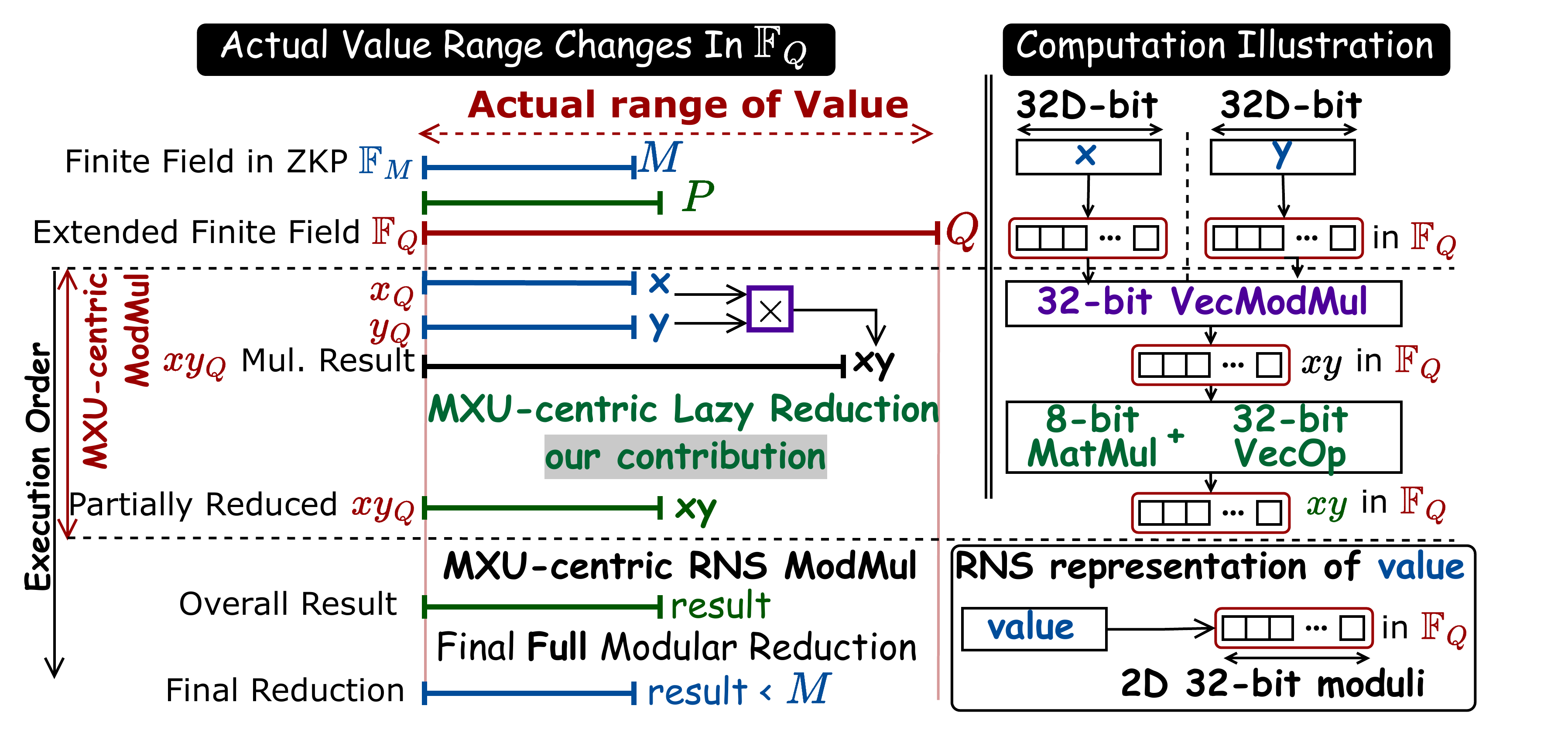}
    \vspace{-8mm}
    \caption{Illustration of actual value change and computational flow of \textit{MXU-centric RNS Lazy Modular Multiplication}.} 
    \label{fig:lazy_reduction}
    \vspace{-5mm}
\end{figure}
Given $x,y\in\Fmxs$, we convert them into extended-field RNS form $x_{\Rqxs},y_{\Rqxs}$, each containing $2D$ 32-bit digits. The modular multiplication proceeds as:  \\
\noindent $\bullet$ RNS-level 32-bit VecMul with Montgomery reduction, $O(2D)$. \\ 
\noindent $\bullet$ MXU-centric RNS Lazy Reduction (Lines 14–20 of Alg.\ref{alg:mxu_centric_modmul}, $O(D^2)$). \\
All sequential carry propagation are removed. Computation becomes dominated by vectorized operations, as highlighted in \tabref{tab:arithmetic_bigt}.

In summary, \textbf{\framework encapsulates computation over $\Fmxs$ inside a larger field $\Fqxs$, paying}
(1) a 2$\times$ memory footprint for the extended RNS representation, and
(2) additional RNS reconstruction/decomposition.
These costs are amortized among sea-of-MACs in MXU, such that
the bottleneck shifts from sequential shuffle-bound carry propagation to VPU-bounded vectorized operations, enabling better throughput/latency than Radix Mont.
\label{sec:rns_lazy_reduction}

\begin{algorithm}[t]
\caption{MXU-Centric RNS Lazy Reduction \textbf{(MXU RNS Lazy)}}
\label{alg:mxu_centric_modmul}
\begin{algorithmic}[1]

\STATE \noindent \hspace{-6mm} \textbf{\textsc{ByteDecompose}}{($x$)}{}$\rightarrow \left[x\right]_{b}, b \in [0,B)$.\\
\STATE \textbf{for} $b = 0$ \textbf{to} $B-1$:
\STATE \quad  $x_{b}\leftarrow $ $x\left[8b: 8(b+1) \right]$ \hfill $\triangleright$ Select b-th byte of $x$, full parallel
\STATE \textbf{Return} $\left[x\right]_{b}, {0\le b <B }$
\vspace{1mm}
\hrule
\vspace{0.6mm}
\noindent \hspace{-6mm} \textbf{\textsc{ByteMerge}{($\left[x\right]_{h}, h \in[0,H)$)}{} $\rightarrow x$}
\STATE \textbf{for} $h = 0$ \textbf{to} $H-1$:
\STATE \quad $x\left[8h: 8(h+1) \right] = \left[x\right]_h$ \hfill $\triangleright$ Parallel among H bytes
\STATE \textbf{Return} $x$ 
\vspace{1mm}
\hrule
\vspace{0.6mm}

\noindent \hspace{-6mm} \textbf{\textsc{Precomputation}}{(${\Rqxs}, {\Gpxs}, w$)}
  \REQUIRE ${\Rqxs}$; ${\Gpxs}$; $w$: the Montgomery factor used in element wise reduction;  $y=(2^{w})^{-1}$, $z = 2^{w}$, $y$ and $z$ are co-prime. 
  \STATE Initialize $I_{i}$, $I_{i,b}$, $\left[E\right]_{i,b,j,h}$, $\left[{f}\right]_i$.
  \STATE \quad $I_{i} = \big(\big((\tfrac{{\Rqxs}}{q_i})^{-1} \bmod p_i\big)(\tfrac{\Rqxs}{q_i}y)\big)\bmod {{\Rqxs}}$
  \STATE $I_{i,b} = \big(\big((\tfrac{{\Rqxs}}{q_i})^{-1} \bmod p_i\big)(\tfrac{\Rqxs}{q_i}y)\big) <<8b\bmod {{\Rqxs}}$
\hfill$\triangleright$ BAT\cite{11408507}
  \STATE $\left[E\right]_{i,b,j,h} = \textsc{ByteDecompose} \big(\big(z(I_{i,b} \bmod{{\Bm}})\big) \bmod p_j\big)_h$

  \STATE $\left[f\right]_i = \Big\lceil \tfrac{I_i 2^u}{{\Rqxs}} \Big\rceil$

  \STATE $\left[g\right]_j = (-z({\Rqxs} \bmod{{\Bm}}))\bmod p_j$
  
  \STATE \textbf{Return} $\left[E\right]_{i,k},\,\left[{f}\right]_i,\,\left[{g}\right]_j$ \\ 
\vspace{1mm}
\hrule
\vspace{0.6mm}

\noindent \hspace{-6mm} \textbf{\textsc{Main}}{(${x}_{{\Rqxs}}$, \Rqxs, \Gpxs, w)}  \hfill$\triangleright$ MXU-centric RNS Lazy Reduction
  \REQUIRE ${\Rqxs}=\prod_i q_i$; ${\Gpxs}=\prod_j p_j$,$i\in[0,I), j\in[0,J)$; w; pick $u\geq \lceil log2(\sum_i q_i) \rceil$. ${x}_{\Rqxs}=\text{CRT}_{\Rqxs}(x)$ is RNS representation of $x$ in $\Fqxs$, containing $I$ residues with $B$ Bytes each. When being written as $\left[{x}_{\Rqxs}\right]_{i}$, each iteration specifies entire 32-bit residue value of $x_{\Rqxs}$. When being written as $\left[{x}_{\Rqxs}\right]_{i,b}$, $b\in[0,B)$, each iteration specify $b$-th byte of $i$-th residue. $\left[{x}_{\Gpxs}\right]_{j,h}$ has $J$ residues with $H$ bytes each. Note: we refer to \cite{SimonRNS} for details of $\text{CRT}_{\Rqxs}$.

  \ENSURE ${x}_{\Gpxs} = ((((x\times y)\bmod {\Bmxs}) \bmod {\Rqxs})\times z)\bmod {\Gp}$ in {\Fpxs}
  \STATE $\left[E\right]_{i,b,j,h},\,\left[{f}\right]_i,\,\left[{g}\right]_j
          \gets \textsc{Precomputation}({\Rqxs},{\Gpxs},w)$
  \STATE $v \gets \left[{x}_{\Rqxs}\right]_i \cdot \left[{f}\right]_i$
          \hfill$\triangleright$ Dot Product, fused into L18 as one MatMul
  \STATE $k \gets \left\lfloor \tfrac{v}{2^u} \right\rfloor$
          \hfill$\triangleright$ 32-bit Vectorized Shift 
  \STATE $\left[{x}_{\Rqxs}\right]_{i,b} =\textsc{ByteDecompose}(\left[{x}_{\Rqxs}\right]_{i} )$
  \STATE $\left[{x}_{\Gpxs}\right]_{j,h} =\sum\left[{x}_{\Rqxs}\right]_{i,b}\times \left[E\right]_{i,b, j,h}$ \hfill$\triangleright$ (Einsum) uint8 MatMul
  \STATE $\left[{x}_{\Gpxs}\right]_{j}=\textsc{ByteMerge}(\left[{x}_{\Gpxs}\right]_{j,h})$
  \STATE \textbf{Return} $\left[{x}_{\Gpxs}\right]_{j} + k\cdot \left[{g}\right]_j$
          \hfill$\triangleright$ 32-bit ScalarVecMul + VecAdd\\
\end{algorithmic}
\end{algorithm}

\vspace{-3mm}
\subsection{Dataflow Optimizations}
\begin{algorithm}
\caption{Layout Stationary Pippenger (\textbf{LS-PPG})} \label{alg:pippenger_implementation}
\begin{algorithmic}[1]
\REQUIRE Initialize $MSM, S_n, P_n, W_{k}, B_{k,j}, W_{k}$ as 0, $j\!\in\![0,2^c-1], k\!\in\![0,K], N'$(max point count per $B_{k,j}$), definition in \figref{fig:msm}. \hfill
\hfill

\hspace{-4mm} \textbf{\texttt{Bucketize}}
\STATE \textbf{sharding for} {$k \in K$} \hfill $\triangleright${Distributed across multi-TPUs}
\STATE \quad Initialize $P'_k \gets \mathcal{O}$ of shape $[2^c, N']$ \hfill $\triangleright$ $2^c$ buckets.
\STATE \quad $\Pi_k \gets \texttt{argsort}(I_k)$ \hfill $\triangleright${$I_k$: $k$-th slice of all scalars $S_n$}
\STATE \quad Gather $P_n$ based on $\Pi_k$, and scatter them into $I_k[n]$-th bucket in $k$-th window ($W_k$).
\hfill

\hspace{-4mm} \textbf{\texttt{Bucket Accumulation (BA)}}
\STATE Initialize bucket $B_{k,j}$ with zero points.
\STATE \textbf{sharding for} $k \in K$ \hfill $\triangleright${Distribute windows across multi-TPUs}
\STATE \quad \textbf{parallel for} $j \in [1,2^c-1]$
\STATE \quad \quad \textbf{for} {$n' \in N'$}
\STATE \quad \quad \quad $B_{k,j} \mathrel{{+}{=}} P^\prime_{k,n',j}$
\hfill

\hspace{-4mm}\textbf{ \texttt{Bucket Reduction (BR)}} \hfill $\triangleright$ Tree-based Accumulation
\STATE Initialize $W_{k,j}\!\gets\!\mathcal{O}$ for all $k\!\in\![0,K-1]$ and $j\!\in\![0,2^c-1]$
\STATE \textbf{for} {$s = c-1$ down to $0$} \hfill $\triangleright${level of tree }
\STATE \quad Let $B^{(L)}_{k,b} \gets B_{k,2b}$ and $B^{(R)}_{k,b} \gets B_{k,2b+1}$ for $b \in [0,2^s-1]$
\STATE \quad Let $W^{(L)}_{k,b}\!\gets\!W_{k,2b}$ and $W^{(R)}_{k,b}\!\gets\!W_{k,2b+1}$ for $b\!\in\![0,2^s-1]$
\STATE \quad \textbf{sharding for} {$b = 0$ to $2^s-1$}
\STATE \quad \quad \textbf{parallel for} {$k = 0$ to $K-1$}
\STATE \quad \quad \quad  $W_{k,b} \gets W^{(L)}_{k,b} + W^{(R)}_{k,b} + B^{(R)}_{k,b}$
\STATE \quad \quad \quad  $B_{k,b} \gets 2 \cdot \left(B^{(L)}_{k,b} + B^{(R)}_{k,b}\right)$
\STATE $W_{k}\gets W_{k,0}$ \hfill $\triangleright$ root of the tree
\hfill

\hspace{-4mm} \textbf{\texttt{Window Merging (WM)}}
\STATE \textbf{for} $k$ in \textbf{($K$, 0, -1)}: \hfill $\triangleright${\textbf{Reverse} Window Index $k$}
\STATE \quad $MSM \mathrel{{\times}{=}} 2^c $ 
\STATE \quad $MSM \mathrel{{+}{=}} W_k$ 
\STATE Return $MSM$.
\end{algorithmic}
\noindent Note 1: Coordinate and bytes dimensions are hidden for simplicity. \\
\noindent Note 2: $\mathcal{O}$ is the zero point $(0,1,1,0)$ in Twisted Edward curves. \\ 
\noindent Note 3: $B_{k,b}^{(L)},B_{k,b}^{(R)},W_{k,b}^{(L)},W_{k,b}^{(R)}$ are introduced for tree reduction. \\
\noindent Note 4: N' is the maximal number of points among all buckets. \\
\end{algorithm}

\begin{figure*}[t!]
    \centering
    \includegraphics[width=0.9\linewidth]{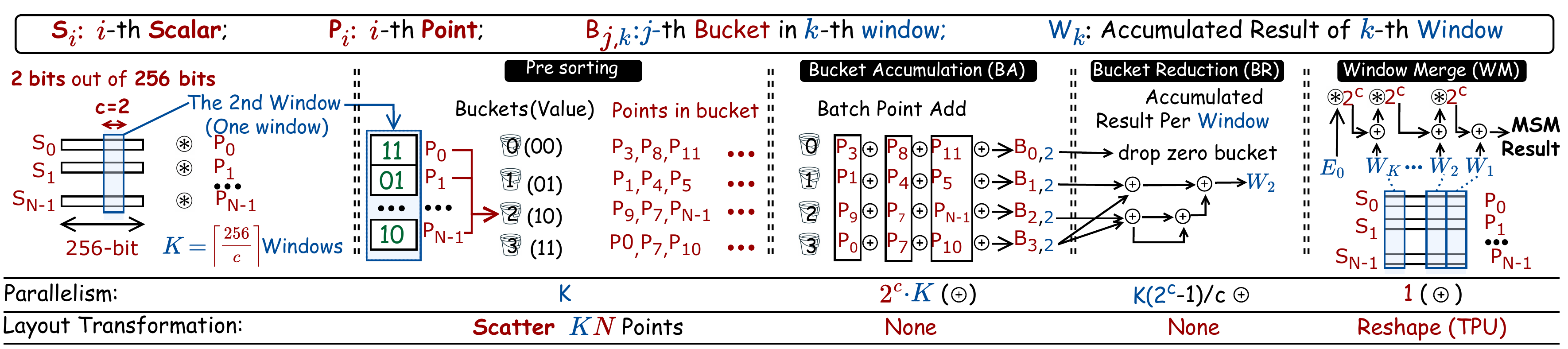}
    \vspace{-4mm}
    \caption{Illustration of LS-PPG's (Alg. \ref{alg:pippenger_implementation}). Takeaway: \framework minimizes data re-sharding and layout reorganization.}
    \label{fig:msm}
    \vspace{-4.3mm}
\end{figure*}

\begin{figure*}[t]
    \setlength{\fboxrule}{5pt}
    \setlength{\fboxsep}{-2pt}
    \centering
    \subfloat[Butterfly NTT \label{fig:ntt_butterfly}]{{\includegraphics[width=0.34\columnwidth]{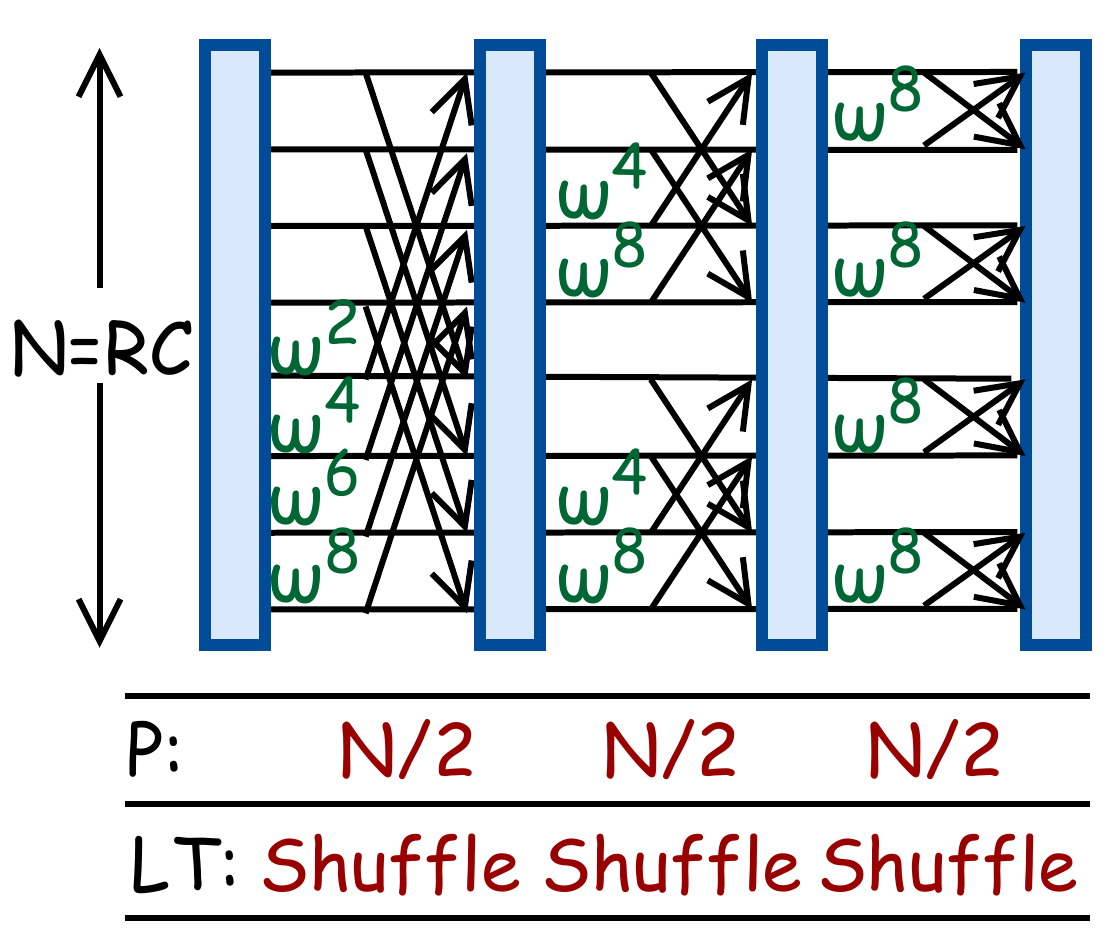}}}
    \subfloat[3-step NTT \label{fig:three_step_ntt}]{{\includegraphics[width=0.64\columnwidth]{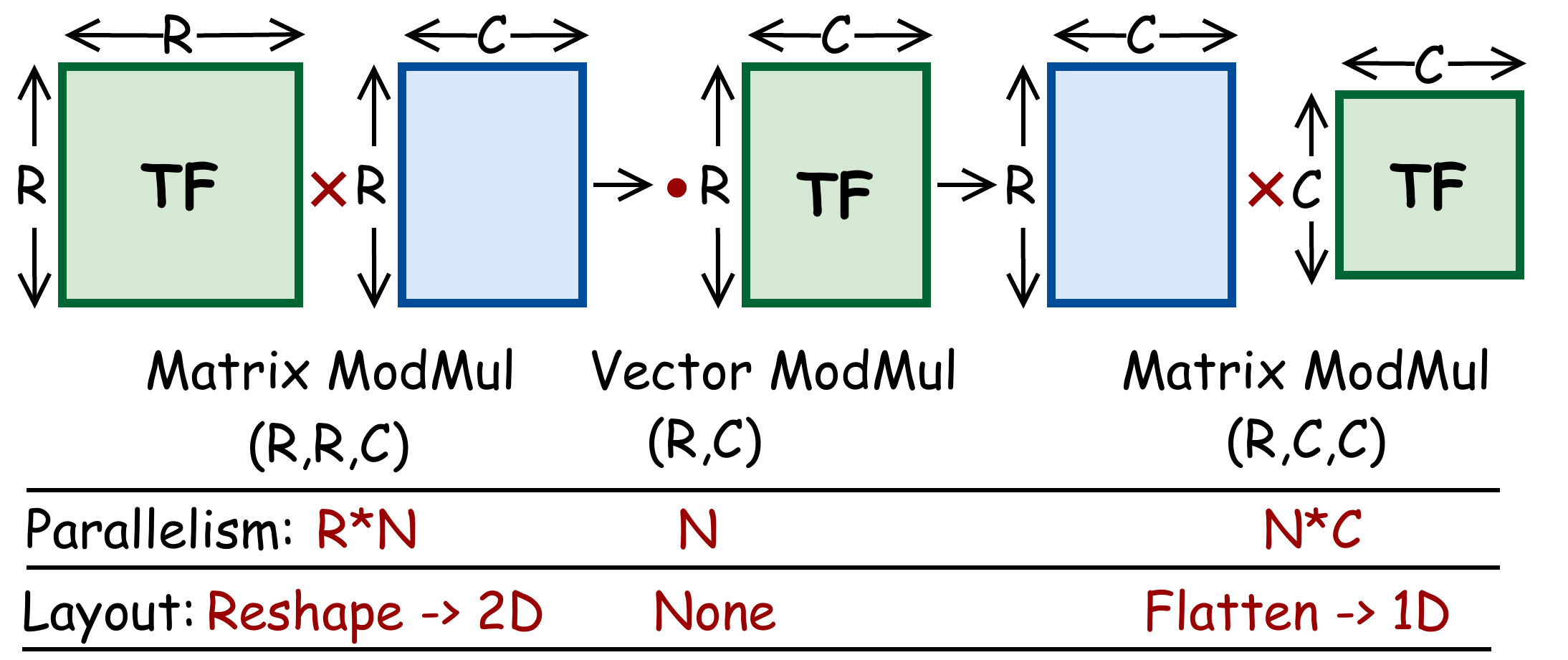}}}
    \subfloat[5-step NTT ({\color{myred}{Proposed by This Work}}) \label{fig:5_step_NTT}]{{\includegraphics[width=1.1\columnwidth]{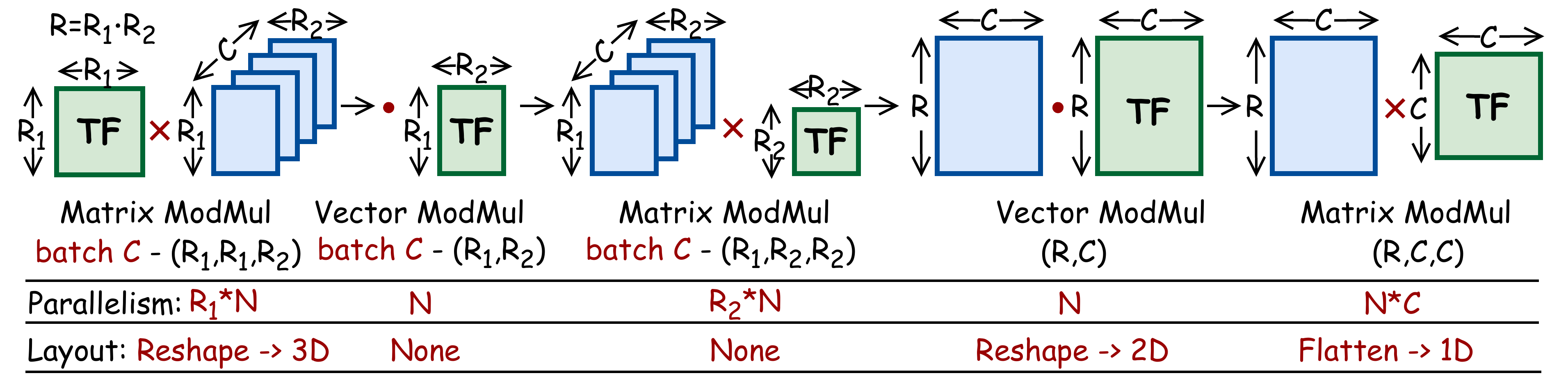}}}
    \vspace{-4mm}
    \caption{Performance Analysis of Different NTT. ($N=R\times C$, $R=R_1\times R_2$), TF refers to twiddle factors, which are generated offline. \textit{Takeaway: \framework further recursively reduces the row-wise NTT in (b) into an 3-step NTT to reduce overall computation. }}
    \vspace{-3mm}
\end{figure*}
\vspace{-3mm}
\subsubsection{Multi-Scalar Multiplication (MSM)} \hfill \\
Challenge. porting SotA MSM to TPU leads to three inefficiencies: (1) \textit{inter-TPU communication}, (2) \textit{intra-TPU layout reorganization}, and (3) \textit{memory overhead} from loading duplicated points of different windows from off-chip memory.



MORPH introduces \emph{Layout-Stationary Pippenger} (\textbf{LS-PPG}), which enforces a common sharding and layout across presorting and Bucket Accumulation (BA), the latency-dominant phase of MSM. LS-PPG shards reduction-free dimensions across tensor cores, allowing each core to process an independent partition. This makes presorting a direct producer of BA-ready points, eliminating cross-TPU communication, intra-TPU layout reorganization. By fusing presorting with BA, post-sorted points are consumed immediately upon generation rather than being written back to and later reloaded from off-chip HBM. Beyond BA, MORPH performs Bucket Reduction in a tree form to expose more parallelism and improve VPU compute utilization. Alg.~\ref{alg:pippenger_implementation} and \figref{fig:msm} illustrate LS-PPG. Consequently, the ideal memory span is reduced from $\frac{KN}{BW_{HBM}}$ to a single pass over scalars and points, i.e., $\frac{2N}{BW_{HBM}}$ in \tabref{tab:ntt_big_t}.

\begin{table}[!htp]\centering
\vspace{-1mm}
\caption{Big-$T$ Complexity of Dataflows (Bottleneck in {\color{myred}{Red}})}\label{tab:ntt_big_t}
\vspace{-4mm}
\setlength{\extrarowheight}{2pt}
\renewcommand{\arraystretch}{1.5}
\resizebox{\columnwidth}{!}{
\begin{tabular}{c|cccc}\hline
\textbf{MSM} &\multicolumn{2}{c}{\textbf{VPU Span} and \textbf{MXU Span}} &\textbf{XLU Span} &\textbf{Memory Span} \\\hline
\rowcolor[gray]{0.9} Presort-PPG &
 \multicolumn{2}{c}{$\displaystyle \frac{KN}{PAR^{BA}_{\text{}}}+\frac{2K(2^c-1)}{PAR^{BR}_{\text{}}} +\frac{(K-1)(1+C)+1}{PAR^{WM}_{\text{}}}$} &
{\color{myred}{$\displaystyle \frac{2^c\cdot N\log N}{PAR_{\text{S}}}$}}  &
{\color{myred}{\textbf{$\displaystyle \frac{K\cdot N}{BW_{\text{HBM}}}$}}}  \\
\rowcolor[HTML]{E6EFDB}LS-PPG &
\multicolumn{2}{c}{$\displaystyle \frac{KN}{PAR^{BA}_{\text{}}}+\frac{4K(2^c-1)}{PAR^{BR\_new}_{\text{}}} +\frac{(K-1)(1+C)+1}{PAR^{WM}_{\text{}}}$}  &
{\color{myred}{$\displaystyle \frac{2^c\cdot N\log N}{PAR_{\text{S}}}$}} &
\textbf{$\displaystyle \frac{2N}{BW_{\text{HBM}}}$} \\
\hline
\textbf{NTT} &\textbf{VPU Span} &\textbf{MXU Span} &\textbf{XLU Span} &\textbf{Memory Span} \\\hline
\rowcolor[gray]{0.9} Butterfly &
$\displaystyle \frac{N\log N}{PAR_{\text{VPU}}}$ &
$\displaystyle 0$ &
{\color{myred}{$\displaystyle \frac{N\log N}{PAR_{\text{S}}}$}} &
\textbf{$\displaystyle \frac{(N+N)}{BW_{\text{HBM}}}$} \\
\rowcolor[gray]{0.9} 3-step NTT &
$\displaystyle \frac{N}{PAR_{\text{VPU}}}$ &
{\color{myred}{\textbf{$\displaystyle \frac{N(R+C)}{PAR_{\text{MXU}}}$}}} &
$\displaystyle \frac{2N}{PAR_{\text{T}}}$ &
$\displaystyle \frac{(2N+R^2+C^2)}{BW_{\text{HBM}}}$ \\
\rowcolor[HTML]{E6EFDB} 5-step NTT  &
$\displaystyle \frac{2N}{PAR_{\text{VPU}}}$ &
{\color{myred}{\textbf{$\displaystyle \frac{N(R_1+R_2+C)}{PAR_{\text{MXU}}}$}}} &
$\displaystyle \frac{3N}{PAR_{\text{T}}}$ &
$\displaystyle \frac{(2N+R_1^2+R_2^2+R+C^2)}{BW_{\text{HBM}}}$ \\ \hline
\end{tabular}}
{\footnotesize Note: $PAR^{BA}$/$PAR^{BR}$/$PAR^{WM}$ refers to $PAR_{VPU}$ or $PAR_{MXU}$ in BA,BR,WM phase. $PAR^{BR\_new}\!=\!c$. $PAR^{BR}\!=\!2$. VPU / MXU runs the same number of PADD.}
\vspace{-5mm}
\end{table}

\subsubsection{Number-Theoretic Transformation (NTT)}   \hfill 

\textit{Challenge of Butterfly NTT and 3-step NTT}.
\tabref{tab:ntt_big_t} shows the Big-$T$ complexity of NTT variants. Running Butterfly NTT on a TPU shows $\frac{PAR_{\text{VPU}}}{PAR_{\text{S}}}\approx O(10^3)$, making XLU the critical bottleneck slowing down butterfly NTT by $O(10^3)$. This is slow even though it offers minimal compute complexity. 
The 3-step NTT~\cite{11408507} removes shuffles by reforming a $N$-degree NTT into two $(R,R,C)$ matrix multiplications and $N$-length vector operations, where $N = R\cdot C$. However, its Big-$T$ is dominated by the MXU span $N(R+C)/PAR_{\text{MXU}}$, which grows with $N$ by a factor of $(R+C)$. Even with balanced factors ($R=C=\sqrt{N}$), it incurs a memory span of at least $4N/BW_{\text{HBM}}$.

\textit{Solution: 5-step NTT to reduce MXU span while preserving MXU utilization.}
To further reduce the MXU and memory span, \framework introduces a \textbf{5-step NTT} that replaces the row-wise $R$-degree NTT (step 1 in the 3-step NTT of \figref{fig:three_step_ntt}) with a 3-step NTT over $(R_1, R_2, C)$, as shown in \figref{fig:5_step_NTT}, where $R = R_1\cdot R_2$. This decomposes the large GEMM into multiple smaller GEMMs, reducing the effective MXU span by a factor of $N^{1/4}$ while being sufficiently large to fully utilize the MXU. The 5-step NTT is detailed in Eq.~\ref{eq:five_ntt}.
\begin{equation}
\label{eq:five_ntt}
TF^{R}_{R\times R} @ \Big[  TF^{C\times R_2}_{R_2\times R_2} @ \Big[ \big( a_{R_1\times C\times R_2} @ TF^{C\cdot R_1}_{R_1\times R_1} \big) \cdot TF^C_{R_1\times R_2} \Big]\cdot TF_{C\times C} \Big]
\end{equation}
Here, \( @ \) denotes matrix multiplication and \(\cdot\) represents element-wise multiplication. $TF^{k}_{R\times C}$ is a matrix $[((\omega_n)^k)^{ij}], i\in[0,R), j\in[0,C)$. $\omega_n$: primitive $N$-th root of unity. $a$ is the input tensor.

\vspace{-3mm}
\section{Experiments}
\label{sec:Experiments}

\subsection{Experiment Setup}

\quad \textbf{Workload:} We take the most popular primitives from zk-SNARKs (degree usually ranges from $2^{14}\sim2^{26}$)~\cite{BenSasson2014, TornadoCash, ZKRollupArchitecture, DarkForest}. We adopt the same evaluation setup as GZKP for MSM and NTT~\cite{gzkp, zprize_msm}, where we assume data comes in affine representation~\cite{short_weierstrass_curves} and offline converted into Twisted Edwards form to minimize compute overhead.

\textbf{TPU Baseline:} We adopt SotA GPU algorithms as our baseline, including (1) radix Montgomery Reduction~\cite{DistMSM,gzkp}, (2) Presorting Pippenger~\cite{DistMSM}, and (3) 3-step NTT ~\cite{11408507}, detailed in \secref{sec:baseline}.

\textbf{\framework TPU Setup:} \framework uses (1) MXU-centric RNS Lazy Reduction, (2) LS-PPG, and (3) 3-step and 5-step NTT. \framework is implemented in JAX~\cite{jax2018github} and captures wall-clock latencies using the JAX Profiler (XProf)~\cite{perfetto}. We use Google TPUv6e as platforms, and scale the number of chips to match V100's power consumption.

\vspace{-2mm}
\subsection{\framework Evaluation}
\begin{figure}[t]
\vspace{-5mm}
    \setlength{\fboxrule}{5pt}
    \setlength{\fboxsep}{-2pt}
    \centering
    \vspace{-2mm}
    {\includegraphics[width=0.413\columnwidth]{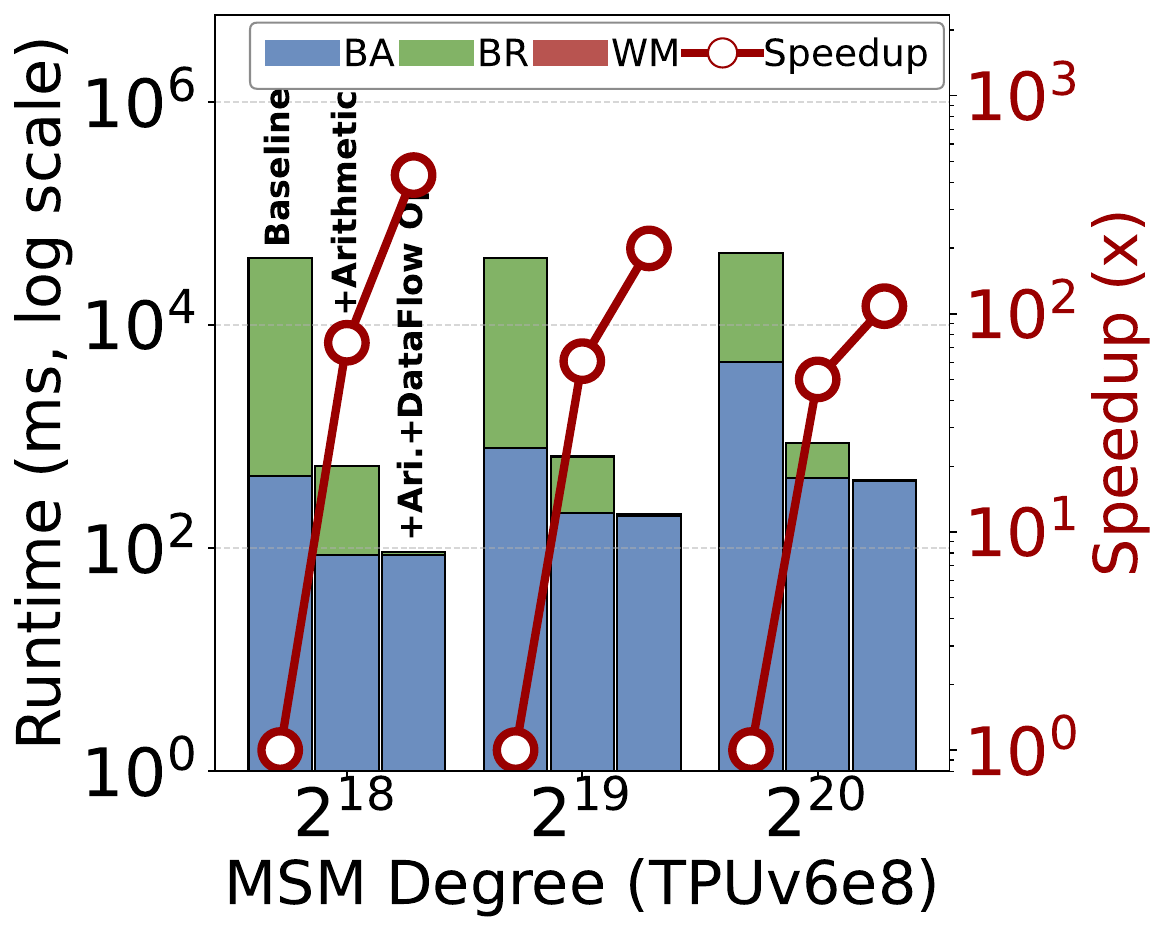}}
    {\includegraphics[width=0.536\columnwidth]{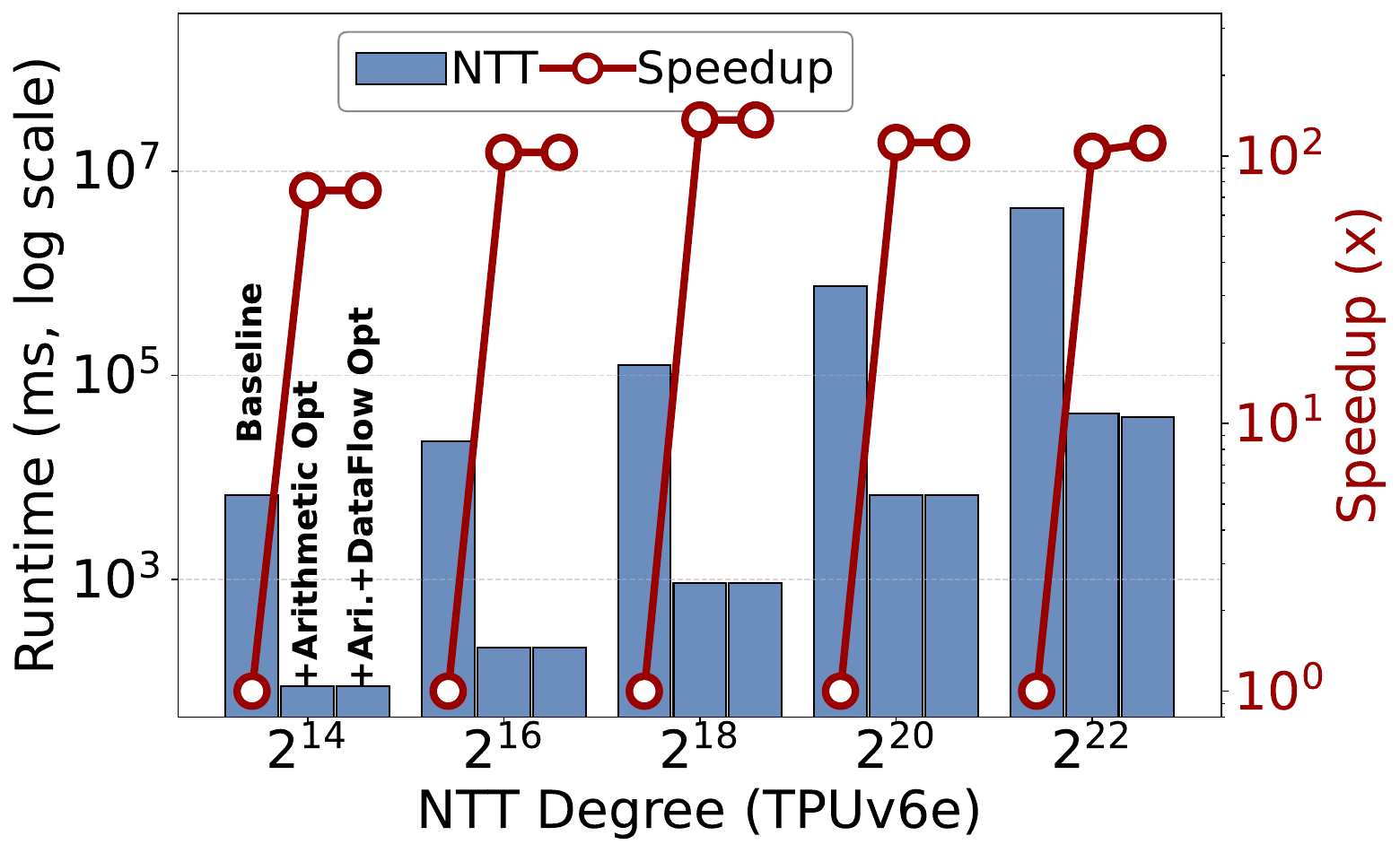}}\\
    \vspace{-5mm} 
    \caption{\framework Ablation Study under different degrees.}
    \label{fig:ablation_study}
    \vspace{-5mm}
\end{figure}

We conduct three-stage ablation for MSM and NTT (Baseline, +Arithmetic, +Ari. + Dataflow, ) with latency and speedup plotted in \figref{fig:ablation_study}.

\vspace{-2mm}
\subsubsection{Arithmetic Optimization Evaluation}
Arithmetic optimizations yield \(\mathbf{50\sim90\times}\) latency reduction for MSM, collapsing BR/WM to near-constant cost. These gains come from \textbf{MXU-centric RNS Lazy Reduction}, which removes sequential carry propagation and reformulating modular reduction of big integer arithmetic into low-precision dense matrix multiplication, allowing the MXU to amortize native \(O(D^2)\) computations into $O(D)$ latency. Although this increases memory footprint by up to \(2\times\), MSM remains compute-bound on VPU operations, so off-chip latency is effectively hidden. Overall, MXU-accelerated RNS-lazy arithmetic is the \textbf{primary contributor} to performance improvement for both MSM and NTT.

\vspace{-2mm}
\subsubsection{MSM -- Dataflow Optimization Evaluation}
\frameworknospace enables up-to 3.1$\times$ speedup. Across all degrees, \textbf{BA (Bucketized included)} dominates the runtime, because of random scatter and gather. Dataflow optimizations reduce BR by $\sim6\times$ from parallelism increase. 

\vspace{-2mm}
\subsubsection{NTT -- Dataflow Optimization Evaluation}
\textbf{3-step NTT is faster at lower degrees}, while \textbf{5-step becomes faster at higher degrees with up-to 1.07$\times$ speedup}. 5-step NTT forces the overall degree $N$ to be decomposed into three factors. At small degree, these factors are typically $\le 2^{6}=64$, leading to poor utilization of the $(8,128)$ VReg shape and reducing utilization for the MXU and VPU. As the degree grows ($>2^{22}$), these factors become large enough to fully utilize VRegs, and 5-step benefits from reduced storage overhead of twiddle parameters. 

\vspace{-2mm}
\subsection{\framework vs. SotAs}
\vspace{-1mm}
\begin{table}[!htp]\centering
\vspace{-8mm}
\caption{Latency Comparison (\framework-tpuv6e8 v.s. GZKP-V100). Unit: $\mu$s/ms for NTT/MSM. Improvement in {\color{myred}{red}}.}
\vspace{-4mm}
\label{tab:morph_vs_gzkp}
\resizebox{0.9\columnwidth}{!}{
\begin{tabular}{c|c|cc|cc}
\hline
\multirow{2}{*}{\textbf{Workload}} 
& \multirow{2}{*}{\textbf{Scale}} 
& \multicolumn{2}{c|}{\textbf{753-bit}} 
& \multicolumn{2}{c}{\textbf{256-bit}} \\
& & \textbf{GZKP} & \textbf{MORPH} & \textbf{GZKP} & \textbf{MORPH} \\
\hline

\multirow{7}{*}{\textbf{NTT} ($\mu$s)}
& $2^{14}$ & 150 & 15.1 & 50 & 11.6 \\
& $2^{16}$ & 490 & 52.8 & 90 & 24.8 \\
& $2^{18}$ & 910 & 337 & 280 & 109 \\
& $2^{20}$ & 7,460 & 2,195 & 1,070 & 716 \\
& $2^{22}$ & 33,670 & 13,107 & 4,960 & 4,694 \\
& $2^{24}$ & 141,400 & OOM & 20,990 & 21,382 \\
\rowcolor[HTML]{E6EFDB}Throughput $\uparrow$  & 
&  & {\color{myred}{\textbf{2.6--10}$\times$}} 
&  & {\color{myred}{\textbf{0.98--4.3}$\times$}} \\
\hline

\multirow{4}{*}{\textbf{MSM} (ms)}
& $N=2^{22}$ & 2.66 & 2.24 & 0.24 & 1.61 \\
& $N=2^{24}$ & 11.30 & 8.91 & 1.10 & 6.42 \\
& $N=2^{26}$ & 40.70 & 35.64 & 4.00 & 25.64 \\
\rowcolor[HTML]{E6EFDB}Throughput $\uparrow$ &  
&  & {\color{myred}{\textbf{1.14--1.2}$\times$}} 
&  & {\color{myred}{\textbf{0.15--0.16}}} \\
\hline
\end{tabular}
}
\vspace{-6mm}
\end{table}

\subsubsection{NTT} \tabref{tab:morph_vs_gzkp} highlights two core advantages of TPUs for large-precision NTTs. 
\textbf{First, TPUv6e8 achieves SoTA throughput in NTT}: Its leading energy efficiency transfers to 10$\times$ higher throughput than GZKP (V100) for 753-bit NTTs. 
\textbf{Second, TPUs scale more gracefully with precision.} Increasing the modulus from 
256- to 753-bits raises GPU latency by $6\times\!~\!\sim7\times$ due to long Montgomery carry chains, 
whereas TPU latency grows only $1.3\times\!\sim3\!\times$ because RNS arithmetic maps efficiently onto the MXU. TPUs also avoid butterfly shuffling entirely. Their cost is determined by the 
compute spans in \tabref{tab:ntt_big_t}, including MatMul 
${(N(R{+}C))}/{PAR_{\text{MXU}}}$, XLU twiddle steps ($2N\sim3N$), and  
${(2N + R^2 + C^2)}/{BW_{\text{HBM}}}$ memory span, all of which are small for lower-degree NTTs.



\subsubsection{MSM}
\tabref{tab:morph_vs_gzkp} reports estimated latency relative to GZKP. \framework’s LS-PPG allows TPUv6e8 to achieve competitive throughput for 753-bit MSM. The latency is dominated by Bucketize, whose per-point scatter exhibits poor memory-bandwidth utilization. This overhead becomes more severe with finer-grained point scatter, leading to lower performance than GZKP at 256-bit. A dedicated hardware reordering unit that hides fine-grained scatters behind computation could mitigate this bandwidth bottleneck~\cite{tong2024FEATHER, tong2026MINISA}.




\begin{figure}[t]
    \setlength{\fboxrule}{5pt}
    \setlength{\fboxsep}{-2pt}
    \centering
    {\includegraphics[width=0.48\columnwidth]{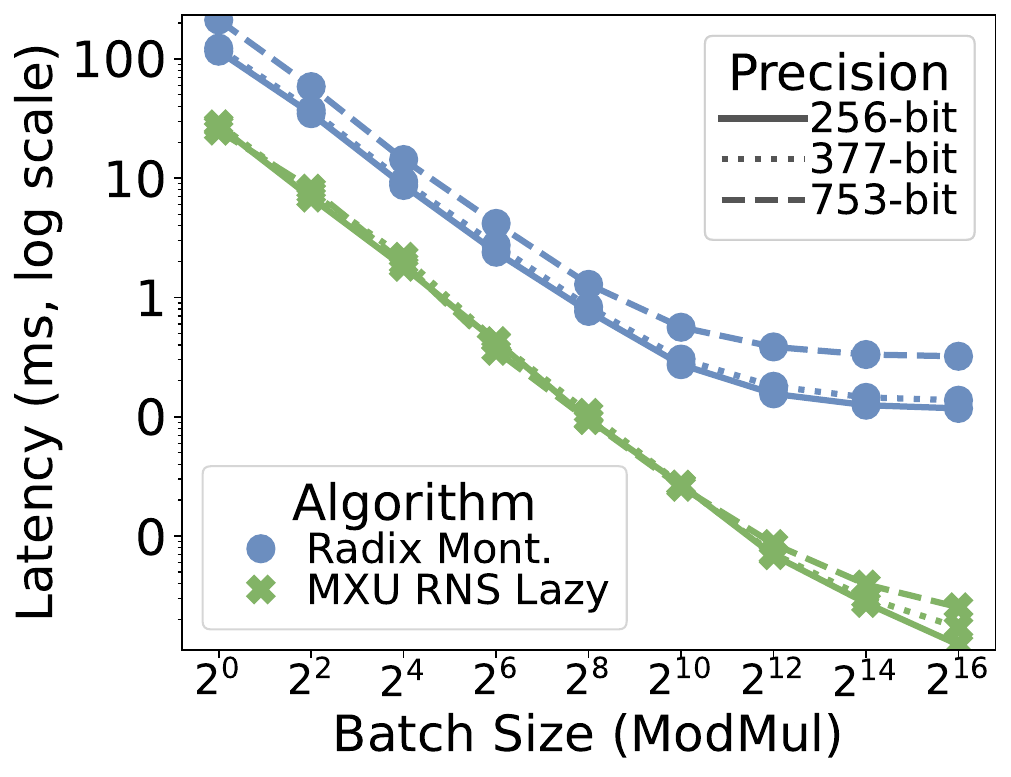}}
    {\includegraphics[width=0.49\columnwidth]{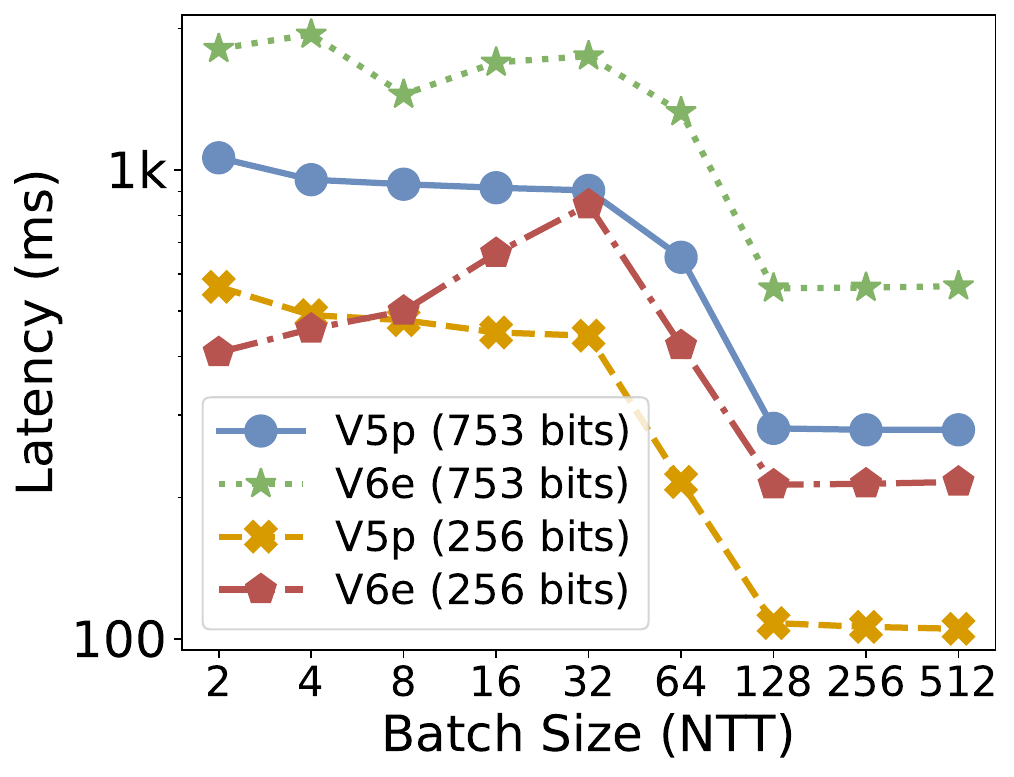}}
    \vspace{-4mm}
    \caption{ModMul and NTT under different batch sizes.}
    \label{fig:ablation_study_batch}
    \vspace{-7.2mm}
\end{figure}



\vspace{-2mm}
\subsection{Ablation Study - Batch Size}
MXU-centric RNS-Lazy sustains \textbf{4$\sim$157$\times$} lower latency than Radix Montgomery across all batch sizes and precisions, and the performance gap widens as precision increases (256$\rightarrow$377$\rightarrow$753 bits) and as batch size increases, it further amplifies MXU utilization (\figref{fig:ablation_study_batch}).

Increasing batch size from 1 to 128 gives \textbf{3.8/3.2$\times$} and \textbf{5.4/3.9$\times$} speedup on NTT (753 and 256 bit) for TPUv5p/v6e (\figref{fig:ablation_study_batch}). \textbf{NTT favors $\!\geq\!128$ batches} to fully utilize VRegs.
Latency plateaus beyond 128 batch size as MXU/VPU achieves its peak compute utilization.
\vspace{-2mm}
\section{Conclusion}
\label{sec:Conclusion}
This paper presents \frameworknospace, the first framework to accelerate ZKP kernels on AI ASICs, deployed on Google TPU. Using a Big-T formulation, we identify the two core bottlenecks and show how to overcome them: bridging high-precision modular arithmetic (>256-bit) to low-precision matrix/vector engines via an \textbf{extended non-prime RNS field}, and eliminating or hiding layout-reorganization and inter-device communication overhead through \textbf{dataflow with layout-invariant sharding over non-reduction dimensions}. \frameworknospace{} makes TPU the SotA commodity platform for high-precision NTT throughput and position AI ASICs as a practical foundation for full ZKP protocol acceleration.

\subsection{Acknowledgments}
This work was supported in part by ACE, one of the seven centers in JUMP 2.0, a Semiconductor Research Corporation (SRC) program sponsored by DARPA. We thank reviewers for their feedbacks.

\bibliographystyle{ACM-Reference-Format}
\bibliography{references}


\begin{thebibliography}{42}


\ifx \showCODEN    \undefined \def \showCODEN     #1{\unskip}     \fi
\ifx \showISBNx    \undefined \def \showISBNx     #1{\unskip}     \fi
\ifx \showISBNxiii \undefined \def \showISBNxiii  #1{\unskip}     \fi
\ifx \showISSN     \undefined \def \showISSN      #1{\unskip}     \fi
\ifx \showLCCN     \undefined \def \showLCCN      #1{\unskip}     \fi
\ifx \shownote     \undefined \def \shownote      #1{#1}          \fi
\ifx \showarticletitle \undefined \def \showarticletitle #1{#1}   \fi
\ifx \showURL      \undefined \def \showURL       {\relax}        \fi
\providecommand\bibfield[2]{#2}
\providecommand\bibinfo[2]{#2}
\providecommand\natexlab[1]{#1}
\providecommand\showeprint[2][]{arXiv:#2}

\bibitem[zpr({[n.\,d.]})]%
        {zprize_msm}
 \bibinfo{year}{[n.\,d.]}\natexlab{}.
\newblock \bibinfo{title}{Benchmark harness for FPGA MSM implementations in the ZPRIZE competition}.
\newblock \bibinfo{howpublished}{\url{https://github.com/z-prize/prize-gpu-fpga-msm/tree/main/harness}}.
\newblock
\newblock
\shownote{Accessed: April 1, 2025}.


\bibitem[sho({[n.\,d.]})]%
        {short_weierstrass_curves}
 \bibinfo{year}{[n.\,d.]}\natexlab{}.
\newblock \bibinfo{title}{XYZZ coordinates for short Weierstrass curves}.
\newblock \bibinfo{howpublished}{\url{https://www.hyperelliptic.org/EFD/g1p/auto-shortw-xyzz.html}}.
\newblock
\newblock
\shownote{Accessed: April 9, 2025}.


\bibitem[ZKR({[n.\,d.]})]%
        {ZKRollupArchitecture}
 \bibinfo{year}{[n.\,d.]}\natexlab{}.
\newblock \bibinfo{title}{ZK Rollup Architecture}.
\newblock \bibinfo{howpublished}{Online}.
\newblock
\urldef\tempurl%
\url{https://zksync.io/faq/tech.html#zk-rollup-architecture}
\showURL{%
\tempurl}
\newblock
\shownote{Accessed: April 2025}.


\bibitem[Tra(2025)]%
        {Trainium}
 \bibinfo{year}{2025}\natexlab{}.
\newblock \bibinfo{title}{Amazon Trainium}.
\newblock \bibinfo{howpublished}{\url{https://aws.amazon.com/ai/machine-learning/trainium/}}.
\newblock
\newblock
\shownote{Accessed: [Insert Date Accessed, e.g., April 9, 2025]}.


\bibitem[per(2025)]%
        {perfetto}
 \bibinfo{year}{2025}\natexlab{}.
\newblock \bibinfo{title}{Perfetto Trace Viewer}.
\newblock \bibinfo{howpublished}{\url{https://perfetto.dev/}}.
\newblock
\newblock
\shownote{Accessed: [Insert Date Accessed, e.g., April 9, 2025]}.


\bibitem[yrr(2025)]%
        {yrrid_gpu_msm}
 \bibinfo{year}{2025}\natexlab{}.
\newblock \bibinfo{title}{yrrid GPU MSM Library}.
\newblock \bibinfo{howpublished}{\url{https://github.com/yrrid/combined-msm-gpu}}.
\newblock
\newblock
\shownote{Accessed: [Insert Date Accessed, e.g., April 9, 2025]}.


\bibitem[Aasaraai et~al\mbox{.}(2022)]%
        {CycloneMSM}
\bibfield{author}{\bibinfo{person}{Kaveh Aasaraai}, \bibinfo{person}{Don Beaver}, \bibinfo{person}{Emanuele Cesena}, \bibinfo{person}{Rahul Maganti}, \bibinfo{person}{Nicolas Stalder}, {and} \bibinfo{person}{Javier Varela}.} \bibinfo{year}{2022}\natexlab{}.
\newblock \bibinfo{booktitle}{\emph{CycloneMSM: FPGA Acceleration of Multi-Scalar Multiplication}}.
\newblock \bibinfo{type}{{T}echnical {R}eport}. \bibinfo{institution}{IACR}.
\newblock
\newblock
\shownote{\url{https://eprint.iacr.org/2022/1396.pdf}}.


\bibitem[Austin et~al\mbox{.}(2025)]%
        {scaling_book}
\bibfield{author}{\bibinfo{person}{Jacob Austin}, \bibinfo{person}{Sholto Douglas}, \bibinfo{person}{Roy Frostig}, \bibinfo{person}{Anselm Levskaya}, \bibinfo{person}{Charlie Chen}, \bibinfo{person}{Sharad Vikram}, \bibinfo{person}{Federico Lebron}, \bibinfo{person}{Peter Choy}, \bibinfo{person}{Vinay Ramasesh}, \bibinfo{person}{Albert Webson}, {and} \bibinfo{person}{Reiner Pope}.} \bibinfo{year}{2025}\natexlab{}.
\newblock \showarticletitle{How to Scale Your Model}.
\newblock \bibinfo{howpublished}{Online}.
\newblock  (\bibinfo{year}{2025}).
\newblock
\newblock
\shownote{Retrieved from https://jax-ml.github.io/scaling-book/}.


\bibitem[Ben-Sasson et~al\mbox{.}(2014)]%
        {BenSasson2014}
\bibfield{author}{\bibinfo{person}{Eli Ben-Sasson}, \bibinfo{person}{Alessandro Chiesa}, \bibinfo{person}{Christina Garman}, \bibinfo{person}{Matthew Green}, \bibinfo{person}{Ian Miers}, \bibinfo{person}{Eran Tromer}, {and} \bibinfo{person}{Madars Virza}.} \bibinfo{year}{2014}\natexlab{}.
\newblock \bibinfo{booktitle}{\emph{Zerocash: Decentralized Anonymous Payments from Bitcoin}}.
\newblock \bibinfo{type}{Technical Report}. \bibinfo{institution}{Zerocash Project}.
\newblock
\urldef\tempurl%
\url{http://zerocash-project.org/media/pdf/zerocash-extended-20140518.pdf}
\showURL{%
\tempurl}


\bibitem[Bradbury et~al\mbox{.}(2018)]%
        {jax2018github}
\bibfield{author}{\bibinfo{person}{James Bradbury}, \bibinfo{person}{Roy Frostig}, \bibinfo{person}{Peter Hawkins}, \bibinfo{person}{Matthew~James Johnson}, \bibinfo{person}{Chris Leary}, \bibinfo{person}{Dougal Maclaurin}, \bibinfo{person}{George Necula}, \bibinfo{person}{Adam Paszke}, \bibinfo{person}{Jake Vander{P}las}, \bibinfo{person}{Skye Wanderman-{M}ilne}, {and} \bibinfo{person}{Qiao Zhang}.} \bibinfo{year}{2018}\natexlab{}.
\newblock \bibinfo{booktitle}{\emph{{JAX}: composable transformations of {P}ython+{N}um{P}y programs}}.
\newblock
\urldef\tempurl%
\url{http://github.com/jax-ml/jax}
\showURL{%
\tempurl}


\bibitem[Buss et~al\mbox{.}(2021)]%
        {intelhexl}
\bibfield{author}{\bibinfo{person}{Jeff Buss} {et~al\mbox{.}}} \bibinfo{year}{2021}\natexlab{}.
\newblock \showarticletitle{Intel HEXL: High-Performance Homomorphic Encryption Primitives}. In \bibinfo{booktitle}{\emph{Proceedings of the Workshop on Encrypted Computing \& Applied Homomorphic Cryptography (WAHC)}}.
\newblock


\bibitem[Choi et~al\mbox{.}(2025)]%
        {kim2024cheddar}
\bibfield{author}{\bibinfo{person}{Wonseok Choi}, \bibinfo{person}{Jongmin Kim}, {and} \bibinfo{person}{Jung~Ho Ahn}.} \bibinfo{year}{2025}\natexlab{}.
\newblock \showarticletitle{Cheddar: A Swift Fully Homomorphic Encryption Library Designed for GPU Architectures}. In \bibinfo{booktitle}{\emph{Proceedings of the 31st ACM International Conference on Architectural Support for Programming Languages and Operating Systems, Volume 1}} (USA) \emph{(\bibinfo{series}{ASPLOS '26})}. \bibinfo{publisher}{Association for Computing Machinery}.
\newblock
\showISBNx{9798400721656}
\href{https://doi.org/10.1145/3760250.3762223}{doi:\nolinkurl{10.1145/3760250.3762223}}


\bibitem[Daftardar et~al\mbox{.}(2024)]%
        {SZKP}
\bibfield{author}{\bibinfo{person}{Alhad Daftardar}, \bibinfo{person}{Brandon Reagen}, {and} \bibinfo{person}{Siddharth Garg}.} \bibinfo{year}{2024}\natexlab{}.
\newblock \showarticletitle{SZKP: A Scalable Accelerator Architecture for Zero-Knowledge Proofs}. In \bibinfo{booktitle}{\emph{Proceedings of the 2024 International Conference on Parallel Architectures and Compilation Techniques}} (Long Beach, CA, USA) \emph{(\bibinfo{series}{PACT '24})}. \bibinfo{publisher}{Association for Computing Machinery}, \bibinfo{address}{New York, NY, USA}, \bibinfo{pages}{271–283}.
\newblock
\showISBNx{9798400706318}
\href{https://doi.org/10.1145/3656019.3676898}{doi:\nolinkurl{10.1145/3656019.3676898}}


\bibitem[Dai and Sunar(2015)]%
        {cuHE}
\bibfield{author}{\bibinfo{person}{Wei Dai} {and} \bibinfo{person}{Berk Sunar}.} \bibinfo{year}{2015}\natexlab{}.
\newblock \showarticletitle{cuHE: A Homomorphic Encryption Accelerator Library}.
\newblock \bibinfo{journal}{\emph{IACR Cryptology ePrint Archive}}  \bibinfo{volume}{2015} (\bibinfo{year}{2015}), \bibinfo{pages}{1043}.
\newblock


\bibitem[{Dark Forest Team}({[n.\,d.]})]%
        {DarkForest}
\bibfield{author}{\bibinfo{person}{{Dark Forest Team}}.} \bibinfo{year}{[n.\,d.]}\natexlab{}.
\newblock \bibinfo{title}{Announcing Dark Forest}.
\newblock \bibinfo{howpublished}{Blog post}.
\newblock
\urldef\tempurl%
\url{https://blog.zkga.me/announcing-darkforest}
\showURL{%
\tempurl}
\newblock
\shownote{Accessed: April 2025}.


\bibitem[Fan et~al\mbox{.}(2022)]%
        {fan2022tensorfheachievingpracticalcomputation}
\bibfield{author}{\bibinfo{person}{Shengyu Fan}, \bibinfo{person}{Zhiwei Wang}, \bibinfo{person}{Weizhi Xu}, \bibinfo{person}{Rui Hou}, \bibinfo{person}{Dan Meng}, {and} \bibinfo{person}{Mingzhe Zhang}.} \bibinfo{year}{2022}\natexlab{}.
\newblock \bibinfo{title}{TensorFHE: Achieving Practical Computation on Encrypted Data Using GPGPU}.
\newblock
\showeprint[arxiv]{2212.14191}~[cs.AR]
\urldef\tempurl%
\url{https://arxiv.org/abs/2212.14191}
\showURL{%
\tempurl}


\bibitem[Gabizon et~al\mbox{.}(2019)]%
        {gabizon2019plonk}
\bibfield{author}{\bibinfo{person}{Ariel Gabizon}, \bibinfo{person}{Zachary~J. Williamson}, {and} \bibinfo{person}{Oana Ciobotaru}.} \bibinfo{year}{2019}\natexlab{}.
\newblock \showarticletitle{Plonk: Permutations over Lagrange-bases for Oecumenical Noninteractive arguments of Knowledge}.
\newblock \bibinfo{journal}{\emph{IACR Cryptol. ePrint Arch.}}  \bibinfo{volume}{2019} (\bibinfo{year}{2019}), \bibinfo{pages}{953}.
\newblock
\newblock
\shownote{https://eprint.iacr.org/2019/953}.


\bibitem[Groth(2016)]%
        {groth2016size}
\bibfield{author}{\bibinfo{person}{Jens Groth}.} \bibinfo{year}{2016}\natexlab{}.
\newblock \showarticletitle{On the Size of Pairing-based Non-interactive Arguments}. In \bibinfo{booktitle}{\emph{Advances in Cryptology - EUROCRYPT 2016}} \emph{(\bibinfo{series}{Lecture Notes in Computer Science}, Vol.~\bibinfo{volume}{9666})}. \bibinfo{publisher}{Springer}, \bibinfo{pages}{305--326}.
\newblock
\href{https://doi.org/10.1007/978-3-662-49896-5_11}{doi:\nolinkurl{10.1007/978-3-662-49896-5_11}}


\bibitem[Jacquemin et~al\mbox{.}(2022)]%
        {large_rns}
\bibfield{author}{\bibinfo{person}{David Jacquemin}, \bibinfo{person}{Ahmet~Can Mert}, {and} \bibinfo{person}{Sujoy~Sinha Roy}.} \bibinfo{year}{2022}\natexlab{}.
\newblock \bibinfo{title}{Exploring {RNS} for Isogeny-based Cryptography}.
\newblock \bibinfo{howpublished}{Cryptology {ePrint} Archive, Paper 2022/1289}.
\newblock


\bibitem[Ji et~al\mbox{.}(2024)]%
        {DistMSM}
\bibfield{author}{\bibinfo{person}{Zhuoran Ji} {et~al\mbox{.}}} \bibinfo{year}{2024}\natexlab{}.
\newblock \showarticletitle{Accelerating Multi-Scalar Multiplication for Efficient Zero Knowledge Proofs with Multi-GPU Systems}. In \bibinfo{booktitle}{\emph{Proceedings of the 29th ACM International Conference on Architectural Support for Programming Languages and Operating Systems, Volume 3}} (La Jolla, CA, USA) \emph{(\bibinfo{series}{ASPLOS '24})}. \bibinfo{publisher}{Association for Computing Machinery}, \bibinfo{address}{New York, NY, USA}, \bibinfo{pages}{57–70}.
\newblock
\showISBNx{9798400703867}
\href{https://doi.org/10.1145/3620666.3651364}{doi:\nolinkurl{10.1145/3620666.3651364}}


\bibitem[Jouppi et~al\mbox{.}(2023)]%
        {jouppi2023tpuv4opticallyreconfigurable}
\bibfield{author}{\bibinfo{person}{Norman~P. Jouppi} {et~al\mbox{.}}} \bibinfo{year}{2023}\natexlab{}.
\newblock \bibinfo{title}{TPU v4: An Optically Reconfigurable Supercomputer for Machine Learning with Hardware Support for Embeddings}.
\newblock
\showeprint[arxiv]{2304.01433}~[cs.AR]
\urldef\tempurl%
\url{https://arxiv.org/abs/2304.01433}
\showURL{%
\tempurl}


\bibitem[Kim et~al\mbox{.}(2023)]%
        {sharp}
\bibfield{author}{\bibinfo{person}{Jongmin Kim}, \bibinfo{person}{Sangpyo Kim}, \bibinfo{person}{Jaewan Choi}, \bibinfo{person}{Jaiyoung Park}, \bibinfo{person}{Donghwan Kim}, {and} \bibinfo{person}{Jung~Ho Ahn}.} \bibinfo{year}{2023}\natexlab{}.
\newblock \showarticletitle{SHARP: A Short-Word Hierarchical Accelerator for Robust and Practical Fully Homomorphic Encryption}. In \bibinfo{booktitle}{\emph{Proceedings of the 50th Annual International Symposium on Computer Architecture}} (Orlando, FL, USA) \emph{(\bibinfo{series}{ISCA '23})}. \bibinfo{publisher}{Association for Computing Machinery}.
\newblock
\showISBNx{9798400700958}
\href{https://doi.org/10.1145/3579371.3589053}{doi:\nolinkurl{10.1145/3579371.3589053}}


\bibitem[Laine et~al\mbox{.}(2018)]%
        {seal}
\bibfield{author}{\bibinfo{person}{Kim Laine}, \bibinfo{person}{Rachel Player}, {and} \bibinfo{person}{Hao Chen}.} \bibinfo{year}{2018}\natexlab{}.
\newblock \showarticletitle{Microsoft SEAL: A Homomorphic Encryption Library}. In \bibinfo{booktitle}{\emph{Proceedings of the IEEE Symposium on Security and Privacy Workshops (SPW)}}. IEEE, \bibinfo{pages}{123--126}.
\newblock


\bibitem[Langowski and Devadas(2025)]%
        {SimonRNS}
\bibfield{author}{\bibinfo{person}{Simon Langowski} {and} \bibinfo{person}{Srinivas Devadas}.} \bibinfo{year}{2025}\natexlab{}.
\newblock \bibinfo{title}{Efficient Modular Multiplication Using Vector Instructions on Commodity Hardware}.
\newblock \bibinfo{howpublished}{Cryptology {ePrint} Archive, Paper 2025/1068}.
\newblock
\urldef\tempurl%
\url{https://eprint.iacr.org/2025/1068}
\showURL{%
\tempurl}


\bibitem[Liu et~al\mbox{.}(2024)]%
        {gypsophila}
\bibfield{author}{\bibinfo{person}{Changxu Liu} {et~al\mbox{.}}} \bibinfo{year}{2024}\natexlab{}.
\newblock \showarticletitle{Gypsophila: A Scalable and Bandwidth-Optimized Multi-Scalar Multiplication Architecture}. In \bibinfo{booktitle}{\emph{Proceedings of the 61st ACM/IEEE Design Automation Conference}} (San Francisco, CA, USA) \emph{(\bibinfo{series}{DAC '24})}. \bibinfo{publisher}{Association for Computing Machinery}, \bibinfo{address}{New York, NY, USA}.
\newblock
\showISBNx{9798400706011}
\href{https://doi.org/10.1145/3649329.3658259}{doi:\nolinkurl{10.1145/3649329.3658259}}


\bibitem[Liu et~al\mbox{.}(2023)]%
        {PriorMSM}
\bibfield{author}{\bibinfo{person}{Changxu Liu}, \bibinfo{person}{Hao Zhou}, \bibinfo{person}{Patrick Dai}, \bibinfo{person}{Li Shang}, {and} \bibinfo{person}{Fan Yang}.} \bibinfo{year}{2023}\natexlab{}.
\newblock \showarticletitle{PriorMSM: An Efficient Acceleration Architecture for Multi-Scalar Multiplication}. In \bibinfo{booktitle}{\emph{Proceedings of the ACM/SIGDA International Symposium on Field-Programmable Gate Arrays}}.
\newblock
\href{https://doi.org/10.1145/3678006}{doi:\nolinkurl{10.1145/3678006}}
\newblock
\shownote{\url{https://dl.acm.org/doi/10.1145/3678006}}.


\bibitem[Ma et~al\mbox{.}(2023b)]%
        {gzkp}
\bibfield{author}{\bibinfo{person}{Tianyu Ma}, \bibinfo{person}{Zhen Zhang}, \bibinfo{person}{Yuhao Zhang}, {and} \bibinfo{person}{G.~Edward Suh}.} \bibinfo{year}{2023}\natexlab{b}.
\newblock \showarticletitle{gZKP: GPU-Accelerated Zero-Knowledge Proof Generation}. In \bibinfo{booktitle}{\emph{Proceedings of the IEEE International Symposium on High-Performance Computer Architecture (HPCA)}}. IEEE.
\newblock


\bibitem[Ma et~al\mbox{.}(2023a)]%
        {ma2023gzkp}
\bibfield{author}{\bibinfo{person}{Weiliang Ma}, \bibinfo{person}{Qian Xiong}, \bibinfo{person}{Xuanhua Shi}, \bibinfo{person}{Xiaosong Ma}, \bibinfo{person}{Hai Jin}, \bibinfo{person}{Haozhao Kuang}, \bibinfo{person}{Mingyu Gao}, \bibinfo{person}{Ye Zhang}, \bibinfo{person}{Haichen Shen}, {and} \bibinfo{person}{Weifang Hu}.} \bibinfo{year}{2023}\natexlab{a}.
\newblock \showarticletitle{GZKP: A GPU Accelerated Zero-Knowledge Proof System}. In \bibinfo{booktitle}{\emph{Proceedings of the 28th ACM International Conference on Architectural Support for Programming Languages and Operating Systems, Volume 2 (ASPLOS 2023)}}. \bibinfo{publisher}{Association for Computing Machinery}, \bibinfo{address}{Vancouver, BC, Canada}, \bibinfo{pages}{340--353}.
\newblock
\href{https://doi.org/10.1145/3575693.3575711}{doi:\nolinkurl{10.1145/3575693.3575711}}


\bibitem[Ohno et~al\mbox{.}(2025)]%
        {ohno2025acceleratingellipticcurvepoint}
\bibfield{author}{\bibinfo{person}{Ayumi Ohno}, \bibinfo{person}{Kotaro Shimamura}, {and} \bibinfo{person}{Shinya Takamaeda-Yamazaki}.} \bibinfo{year}{2025}\natexlab{}.
\newblock \bibinfo{title}{Accelerating Elliptic Curve Point Additions on Versal AI Engine for Multi-scalar Multiplication}.
\newblock
\showeprint[arxiv]{2502.11660}~[cs.AR]
\urldef\tempurl%
\url{https://arxiv.org/abs/2502.11660}
\showURL{%
\tempurl}


\bibitem[Pottier et~al\mbox{.}(2025)]%
        {pottier2025optimsm}
\bibfield{author}{\bibinfo{person}{Xander Pottier}, \bibinfo{person}{Thomas de Ruijter}, \bibinfo{person}{Jonas Bertels}, \bibinfo{person}{Wouter Legiest}, \bibinfo{person}{Michiel Van~Beirendonck}, {and} \bibinfo{person}{Ingrid Verbauwhede}.} \bibinfo{year}{2025}\natexlab{}.
\newblock \showarticletitle{OPTIMSM: FPGA hardware accelerator for Zero-Knowledge MSM}.
\newblock \bibinfo{journal}{\emph{IACR Transactions on Cryptographic Hardware and Embedded Systems}} \bibinfo{volume}{2025}, \bibinfo{number}{2} (\bibinfo{year}{2025}), \bibinfo{pages}{489--510}.
\newblock


\bibitem[Ray et~al\mbox{.}(2023)]%
        {HardcamlMSM}
\bibfield{author}{\bibinfo{person}{Andy Ray}, \bibinfo{person}{Benjamin Devlin}, \bibinfo{person}{Fu~Yong Quah}, {and} \bibinfo{person}{Rahul Yesantharao}.} \bibinfo{year}{2023}\natexlab{}.
\newblock \showarticletitle{Hardcaml MSM: A High-Performance Split CPU-FPGA Multi-Scalar Multiplication Engine}. In \bibinfo{booktitle}{\emph{Proceedings of the ACM Symposium on Field-Programmable Gate Arrays}}.
\newblock
\href{https://doi.org/10.1145/3626202.3637577}{doi:\nolinkurl{10.1145/3626202.3637577}}
\newblock
\shownote{\url{https://dl.acm.org/doi/10.1145/3626202.3637577}}.


\bibitem[Reagen et~al\mbox{.}(2021)]%
        {heax}
\bibfield{author}{\bibinfo{person}{Brandon Reagen}, \bibinfo{person}{Woojoo Choi}, \bibinfo{person}{David Brooks}, \bibinfo{person}{Gu-Yeon Wei}, {and} \bibinfo{person}{Hsien-Hsin~S. Lee}.} \bibinfo{year}{2021}\natexlab{}.
\newblock \showarticletitle{HEAX: An Architecture for Computing on Encrypted Data}. In \bibinfo{booktitle}{\emph{Proceedings of the ACM/IEEE International Symposium on Computer Architecture (ISCA)}}. IEEE, \bibinfo{pages}{1113--1126}.
\newblock


\bibitem[Samardzic et~al\mbox{.}(2024)]%
        {nocap}
\bibfield{author}{\bibinfo{person}{Nikola Samardzic}, \bibinfo{person}{Simon Langowski}, \bibinfo{person}{Srinivas Devadas}, {and} \bibinfo{person}{Daniel Sanchez}.} \bibinfo{year}{2024}\natexlab{}.
\newblock \showarticletitle{Accelerating Zero-Knowledge Proofs Through Hardware-Algorithm Co-Design}. In \bibinfo{booktitle}{\emph{2024 57th IEEE/ACM International Symposium on Microarchitecture (MICRO)}}. \bibinfo{pages}{366--379}.
\newblock
\href{https://doi.org/10.1109/MICRO61859.2024.00035}{doi:\nolinkurl{10.1109/MICRO61859.2024.00035}}


\bibitem[Storm et~al\mbox{.}(2020)]%
        {TornadoCash}
\bibfield{author}{\bibinfo{person}{Roman Storm}, \bibinfo{person}{Alexey Pertsev}, {and} \bibinfo{person}{Roman Semenov}.} \bibinfo{year}{2020}\natexlab{}.
\newblock \bibinfo{title}{Tornado Cash Privacy Solution}.
\newblock
\urldef\tempurl%
\url{https://tornado.cash/Tornado.cash_whitepaper_v1.4.pdf}
\showURL{%
\tempurl}
\newblock
\shownote{White Paper}.


\bibitem[Tong et~al\mbox{.}(2026a)]%
        {11408507}
\bibfield{author}{\bibinfo{person}{Jianming Tong}, \bibinfo{person}{Tianhao Huang}, \bibinfo{person}{Jingtian Dang}, \bibinfo{person}{Leo de Castro}, \bibinfo{person}{Anirudh Itagi}, \bibinfo{person}{Anupam Golder}, \bibinfo{person}{Asra Ali}, \bibinfo{person}{Jeremy Kun}, \bibinfo{person}{Jevin Jiang}, \bibinfo{person}{Arvind}, \bibinfo{person}{G. Edward~Suh}, {and} \bibinfo{person}{Tushar Krishna}.} \bibinfo{year}{2026}\natexlab{a}.
\newblock \showarticletitle{Leveraging ASIC AI Chips for Homomorphic Encryption}. In \bibinfo{booktitle}{\emph{2026 IEEE International Symposium on High Performance Computer Architecture (HPCA)}}. \bibinfo{pages}{1--18}.
\newblock
\href{https://doi.org/10.1109/HPCA68181.2026.11408507}{doi:\nolinkurl{10.1109/HPCA68181.2026.11408507}}


\bibitem[Tong et~al\mbox{.}(2024)]%
        {tong2024FEATHER}
\bibfield{author}{\bibinfo{person}{Jianming Tong}, \bibinfo{person}{Anirudh Itagi}, \bibinfo{person}{Parsanth Chatarasi}, {and} \bibinfo{person}{Tushar Krishna}.} \bibinfo{year}{2024}\natexlab{}.
\newblock \showarticletitle{FEATHER: A Reconfigurable Accelerator with Data Reordering Support for Low-Cost On-Chip Dataflow Switching}. In \bibinfo{booktitle}{\emph{Proceedings of the 51th Annual International Symposium on Computer Architecture}} (Argentina) \emph{(\bibinfo{series}{ISCA '24})}. \bibinfo{publisher}{Association for Computing Machinery}, \bibinfo{address}{Argentina}.
\newblock


\bibitem[Tong et~al\mbox{.}(2026b)]%
        {tong2026MINISA}
\bibfield{author}{\bibinfo{person}{Jianming Tong}, \bibinfo{person}{Yujie Li}, \bibinfo{person}{Devansh Jain}, \bibinfo{person}{Charith Mendis}, {and} \bibinfo{person}{Tushar Krishna}.} \bibinfo{year}{2026}\natexlab{b}.
\newblock \showarticletitle{MINISA: Minimal Instruction Set Architecture for Next-gen Reconfigurable Inference Accelerator}. In \bibinfo{booktitle}{\emph{Proceedings of the 34th Annual International Symposium on Performance Analysis of Systems and Software}} (Seoul, Korea) \emph{(\bibinfo{series}{ISPASS '26})}.
\newblock


\bibitem[Wang and Gao(2025)]%
        {UniZK}
\bibfield{author}{\bibinfo{person}{Cheng Wang} {and} \bibinfo{person}{Mingyu Gao}.} \bibinfo{year}{2025}\natexlab{}.
\newblock \showarticletitle{UniZK: Accelerating Zero-Knowledge Proof with Unified Hardware and Flexible Kernel Mapping}. In \bibinfo{booktitle}{\emph{Proceedings of the 30th ACM International Conference on Architectural Support for Programming Languages and Operating Systems, Volume 1}} (Rotterdam, Netherlands) \emph{(\bibinfo{series}{ASPLOS '25})}. \bibinfo{publisher}{Association for Computing Machinery}, \bibinfo{address}{New York, NY, USA}, \bibinfo{pages}{1101–1117}.
\newblock
\showISBNx{9798400706981}
\href{https://doi.org/10.1145/3669940.3707228}{doi:\nolinkurl{10.1145/3669940.3707228}}


\bibitem[Williams et~al\mbox{.}(2009)]%
        {roofline_model}
\bibfield{author}{\bibinfo{person}{Samuel Williams}, \bibinfo{person}{Andrew Waterman}, {and} \bibinfo{person}{David Patterson}.} \bibinfo{year}{2009}\natexlab{}.
\newblock \showarticletitle{Roofline: an insightful visual performance model for multicore architectures}.
\newblock \bibinfo{journal}{\emph{Commun. ACM}} \bibinfo{volume}{52}, \bibinfo{number}{4} (\bibinfo{date}{April} \bibinfo{year}{2009}), \bibinfo{pages}{65–76}.
\newblock
\showISSN{0001-0782}
\href{https://doi.org/10.1145/1498765.1498785}{doi:\nolinkurl{10.1145/1498765.1498785}}


\bibitem[Yang et~al\mbox{.}(2025)]%
        {legoZK}
\bibfield{author}{\bibinfo{person}{Zhengbang Yang} {et~al\mbox{.}}} \bibinfo{year}{2025}\natexlab{}.
\newblock \showarticletitle{LegoZK: A Dynamically Reconfigurable Accelerator for Zero-Knowledge Proof}. In \bibinfo{booktitle}{\emph{2025 IEEE International Symposium on High Performance Computer Architecture (HPCA)}}.
\newblock
\href{https://doi.org/10.1109/HPCA61900.2025.00020}{doi:\nolinkurl{10.1109/HPCA61900.2025.00020}}


\bibitem[Zhang et~al\mbox{.}(2025)]%
        {zhang2025zkvc}
\bibfield{author}{\bibinfo{person}{Yancheng Zhang} {et~al\mbox{.}}} \bibinfo{year}{2025}\natexlab{}.
\newblock \showarticletitle{zkVC: Fast Zero-Knowledge Proof for Private and Verifiable Computing}. In \bibinfo{booktitle}{\emph{Proceedings of the 62nd Annual ACM/IEEE Design Automation Conference}} (San Francisco, California, United States) \emph{(\bibinfo{series}{DAC '25})}. \bibinfo{publisher}{IEEE Press}.
\newblock
\showISBNx{9798331503048}
\href{https://doi.org/10.1109/DAC63849.2025.11132681}{doi:\nolinkurl{10.1109/DAC63849.2025.11132681}}


\bibitem[Zhang et~al\mbox{.}(2021)]%
        {zhang2021pipezk}
\bibfield{author}{\bibinfo{person}{Ye Zhang}, \bibinfo{person}{Shuo Wang}, \bibinfo{person}{Xian Zhang}, {et~al\mbox{.}}} \bibinfo{year}{2021}\natexlab{}.
\newblock \showarticletitle{PipeZK: Accelerating Zero-Knowledge Proof with a Pipelined Architecture}. In \bibinfo{booktitle}{\emph{Proceedings of the 2021 ACM/IEEE 48th Annual International Symposium on Computer Architecture (ISCA)}}.
\newblock


\end{thebibliography}

\end{document}